\documentclass[amsmath,amssymb,amsfonts,pre,reprint,superscriptaddress,bibnotes,showpacs,showkeys,longbibliography]{revtex4-1}

\usepackage{graphicx}
\usepackage{subfigure}
\usepackage{amsmath}
\usepackage{amsfonts}
\usepackage{amssymb}
\usepackage{natbib}
\usepackage{relsize}
\usepackage{sidecap}

\begin{document}


\title{Generalized Potentials for a Mean-field Density Functional Theory of a Three-Phase Contact Line}


\author{Chang-You Lin}
\email{changyoul@gmail.com}
\thanks{Corresponding author}
\affiliation{Instituut voor Theoretische Fysica, KU Leuven, Celestijnenlaan 200 D, B-3001 Leuven, Belgium}
\affiliation{Department of Physics, Carnegie Mellon University, Pittsburgh, PA 15232, USA}
\author{Michael Widom}
\email{widom@andrew.cmu.edu}
\author{Robert F. Sekerka}
\email{sekerka@cmu.edu}
\affiliation{Department of Physics, Carnegie Mellon University, Pittsburgh, PA 15232, USA}


\date{\today}

\begin{abstract}

We investigate generalized potentials for a mean-field density functional theory of a three-phase contact line. Compared to the symmetrical potential introduced in our previous article \cite{lin2012mean}, the three minima of these potentials form a small triangle located arbitrarily within the Gibbs triangle, which is more realistic for ternary fluid systems. We multiply linear functions that vanish at edges and vertices of the small triangle, yielding potentials in the form of quartic polynomials. We find that a subset of such potentials has simple analytic far-field solutions, and is a linear transformation of our original potential. By scaling, we can relate their solutions to those of our original potential. For special cases, the lengths of the sides of the small triangle are proportional to the corresponding interfacial tensions. For the case of equal interfacial tensions, we calculate a line tension that is proportional to the area of the small triangle.
\end{abstract}

\pacs{05.70.Np, 65.40.gp, 68.05.-n, 68.35.Md}
\keywords{Line tension; interfacial tension; three-phase contact line; mean-field density functional theory; quartic potential; Gibbs triangle}

\maketitle

\section{Introduction}

In our previous article \cite{lin2012mean}, we treated a three-phase contact line by a mean-field density functional that involves a symmetric potential. This potential allows us to find analytical far-field solutions because one of the mole fractions is constant for any two-phase transition. In order to develop a phenomenological model for a non-symmetric potential, which is more realistic for ternary fluid systems, we need a more general potential that has three minima in arbitrary locations within the Gibbs triangle. In this article, we introduce a flexible approach to construct such generalized potentials.

\begin{figure}[h!]
\centering
\subfigure{
\label{subfig:realspace}}
\subfigure{
\label{subfig:gibbsspace}}
\subfigure{
\label{subfig:sliceinterface}}
\includegraphics[scale=0.44]{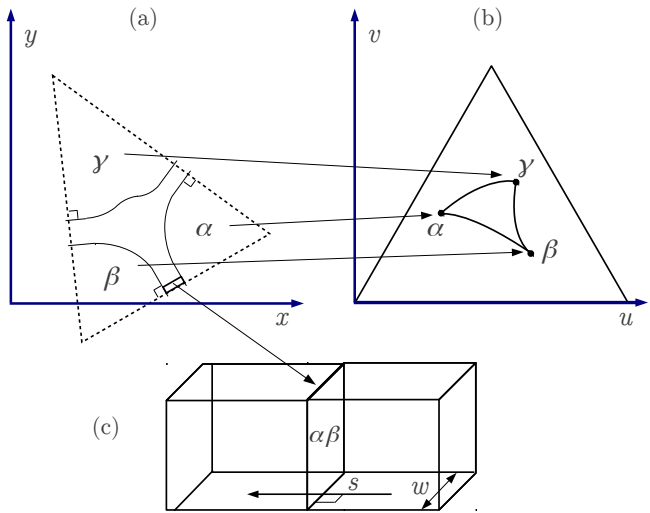}
\caption{\label{fig:realgibbscompare} Schematic diagram for the mapping between physical space and the Gibbs triangle (briefly, the Gibbs space). \subref{subfig:realspace} Sketch of our system consisting of three bulk phases $\alpha$, $\beta$, and $\gamma$, divided by three interfaces $\alpha \beta$, $\beta \gamma$ and $\gamma \alpha$. The interfaces, if extrapolated, are each perpendicular to sides of the dashed triangle (illustrative computational domain) and meet at a three-phase contact line. The corresponding dihedral angles are $\theta_{\alpha}$, $\theta_{\beta}$, and $\theta_{\gamma}$ and $(x,y)$ are Cartesian coordinates in physical space. The region within three solid curves represents approximately the diffuse region of the interfaces and the contact line. \subref{subfig:gibbsspace} The Gibbs space, in which the three bulk phases are located at three points and $(u,v)$ are Cartesian coordinates in units of mole fractions. The three curved lines that connect pairs of points for bulk phases are trajectories of two-phase transitions. \subref{subfig:sliceinterface} Slice of an interface in a cuboid far from the contact line.  $s$  is a coordinate perpendicular to the $\alpha\beta$-interface and $w$ is the width of the cuboid in a direction perpendicular to the contact line.}
\end{figure}

As illustrated geometrically in Fig.~\ref{subfig:realspace}, a three-phase contact line is modeled for a ternary fluid system having three bulk phases $\alpha$, $\beta$, and $\gamma$, which subtend dihedral angles $\theta_{\alpha}$, $\theta_{\beta}$, and $\theta_{\gamma}$ as sketched in Fig.~\ref{subfig:realspace}. Each of the interfaces $\alpha \beta$, $\beta \gamma$ and $\gamma \alpha$, if extrapolated far from the contact line, is perpendicular to a side of the dashed triangle, which is illustrative of our much larger actual computational domain. The contact line is where the three interfaces appear to meet. We assume the system to be translationally invariant along the vertical direction perpendicular to this figure, thus reducing this problem to two dimensions. Compared to the homogeneous bulk phases, the inhomogeneity arises from the formation of interfaces and a contact line, which are actually diffuse regions.

Based on the thermodynamic method introduced by Gibbs \cite[p.~228]{gibbs1928collected}, the inhomogeneity is treated by means of an excess grand potential $\Omega_{xs}$, which, by convention \cite[ch.~8]{rowlinson2002molecular}, can be expressed by
\begin{equation}\label{eq:omegaxsconvention}
\Omega_{xs} = L \tau + L R_{\alpha \beta} \sigma_{\alpha \beta} + L R_{\beta \gamma} \sigma_{\beta \gamma} + L R_{\gamma \alpha} \sigma_{\gamma \alpha}.
\end{equation}
Here, $L$ is the length of the contact line and $R_{ij}$ is the distance from the contact line along interface $ij$ toward the boundary. In the limit of all $R_{ij}\rightarrow \infty$, the line tension $\tau$ is defined as the excess grand potential per unit length associated with the contact line while each of the three interfacial tensions $\sigma_{ij}$ (excess grand potentials per unit area) is associated with an interface $ij$ in the far-field limit. A classical result \cite{neumann1894} shows that  
\begin{equation}\label{eq:eqangle}
\frac{\sin \theta_{\alpha }}{\sigma_{\beta \gamma}} = \frac{\sin \theta_{\beta}}{\sigma_{\alpha  \gamma}} = \frac{\sin \theta_{\gamma}}{\sigma_{\alpha \beta}}.
\end{equation}
According to this result, the boundary is actually a Neumann triangle since the three sides are proportional to three interfacial tensions.

Besides extensive studies of interfacial tensions (See \cite{adamson1997physical}), the line tension of a contact line plays a crucial role in a broad range of physical phenomena such as the equilibrium shapes of small droplets \cite{gaydos1987dependence,drelich1993effect}, microfluidics \cite{grunze1999driven,weigl1999microfluidic}, heterogeneous nucleation \cite{hienola2007estimation}, cell adhesion \cite{sackmann2002cell}, the dynamics of a drop spreading on a liquid thin film \cite{fukai1995wetting,bonn2009wetting} and the behavior of line tension at wetting transitions \cite{indekeu2010wetting}. For a review for both experimental and theoretical aspects of line tension, see \cite{amirfazli2004status}, and for a conceptual review, see \cite{schimmele2007conceptual}.

\subsection{Mean-field Density Functional Theory}\label{sec:mfdft}

According to the mean-field density functional theory introduced in our earlier work \cite{lin2012mean}, the excess grand potential for a ternary fluid system is
\begin{equation}\label{eq:omegaxsgeneral}
\Omega_{xs} = B L \int_A \left[ f +  g \right] \mathrm{d}A,
\end{equation}
where $f$ is a potential function and $g$ is a gradient energy of the chemical constituents. In this article, we assumed that the dominant intermolecular forces in our system are short range, so physical quantities can be formulated in terms of local densities. Based on the assumption of uniform molar volume, we change the variables of $f$ and $g$ from number densities to mole fractions $X_i$, and $\sum_{i=1}^{3} X_i=1$ (see \cite[sec.~II]{lin2012mean} for details). For reviews of the general mean-field density functional method of interfaces, see \cite{rowlinson2002molecular,anderson1998diffuse}.

In this phenomenological theory, there is no definite form of the potential $f$ for the excess grand potential $\Omega_{xs}$ in \eqref{eq:omegaxsgeneral}. The specific form that we used in earlier work \cite{lin2012mean} is
\begin{equation}\label{eq:potential0}
f=\sum\limits_{i=1}^{3}(X_i-a)^2(X_i-b)^2 \equiv f^o,
\end{equation}
where $0 \leq a \leq 1$ is a constant and $b=(1-a)/2$, is a symmetric quartic function that allowed us to obtain asymptotic analytical solutions in the far-field. This form is an extension of the potential used in the Landau's phenomenological theory \cite{LandauLifshitz1935,umantsev2012field}. Similar two-density quartic potentials have been used in \cite{szleifer1992surface,koga2010first} for line tension and for a first order wetting transition. Sixth order polynomials have been used for second and higher order transitions \cite{koga2008mean,koga2010first,koga2010infinite}. In this article, we focus on potentials of quartic form. 

Compared to other two-density models, our model is actually pseudo-binary since it formulates the potential in terms of three mole fractions $\{X_i\}_{i=1,2,3}$ by the assumption of uniform molar volume. Because of the constraint $\sum_{i=1}^{3} X_i=1$, there are two independent mole fractions, so our potential can be described in terms of two independent variables, although we sometimes display all three mole factions to illustrate its symmetry.

Note that the gradient energy $g$ in the functional \eqref{eq:omegaxsgeneral} of the excess grand potential $\Omega_{xs}$ is specified by
\begin{equation}
g=\sum\limits_{i=1}^{3} \frac{\ell_i^2}{2} \left|  \nabla X_i \right|^2,
\end{equation}
where $\{\ell_i\}_{i=1,2,3}$ are constants. For cases with equal $\ell_i$ (isotropic gradient energy), the three-fold symmetric potential leads to three-fold symmetry of the physical domain. To resolve this special geometry, we employed a triangular grid to obtain numerical solutions over the entire domain (for details, see \cite[sec.~III]{lin2012mean}).

\subsection{Mapping from Physical Space to the Gibbs Space}\label{sec:mapping}

In our mean-field density functional theory of a three-phase contact line for a ternary fluid system (summarized in Sec.~\ref{sec:mfdft}; for details, see \cite[sec.~II]{lin2012mean} ), the functional of the excess grand potential consists of a potential function and a gradient energy. Since we use three mole fractions as variables, there is a connection between the physical space and the space of the Gibbs triangle (briefly, the Gibbs space). This connection depends on the specific choice of potential and the coefficients of the gradient energy. Fig.~\ref{fig:realgibbscompare} illustrates a mapping between the physical space of our system \ref{subfig:realspace} and the Gibbs space \ref{subfig:gibbsspace}. In general, the three bulk phases of the physical system are represented by the three minima of a given potential, which are located at three points within the Gibbs triangle.

The curved lines within the Gibbs triangle (Fig.~\ref{subfig:gibbsspace}) are the trajectories of the two phase-transitions in the far-field regime between any pair of bulk phases in the physical space (Fig.~\ref{subfig:realspace}). The far-field regime is located at a distance that is far from the contact line compared to the interfacial widths (or the size of the central core region associated to the contact line). In the far-field regime, a slice of interface can be contained in a cuboid (Fig.~\ref{subfig:sliceinterface}) and the interfacial width is constant. For a two-phase transition between any pair of bulk phases, the change of mole fractions from one bulk phase to another maps to a curved line connecting two potential minima within the Gibbs triangle. At equilibrium, these trajectories minimize the excess grand potential according to the form of the potential and the coefficients of the gradient energy. Since the interfacial tensions are excess grand potentials per unit area, their values depend on these trajectories.

According to the form \eqref{eq:omegaxsconvention} of the excess grand potential, the line tension is the residue of excess potential in which we subtract the contribution from the far-field interfacial tensions. In physical space (Fig.~\ref{subfig:realspace}), it relates to the core of the diffuse region centered at the contact line as if we subtract the cuboids of the interfaces extended from the far-field until they meet at the ``contact line''. This core region corresponds to three-phase transitions among all three bulk phases in the physical space. It can be mapped to a region within the Gibbs triangle (Fig.~\ref{subfig:gibbsspace}) surrounded by the three trajectories of the two-phase transitions. Similar to the interfacial tensions, the line tension is the excess grand potential per unit length. Its value depends on the form of the potential within this core area and, of course, the coefficients of gradient energy.

In this article, we extend our previous model to more realistic systems. Specifically, we use geometrical reasoning to construct generalized quartic potentials with three minima arbitrarily located within the Gibbs triangle. For a subset of these potentials, the resulting potential is a linear transformation of our original potential and we can obtain simple analytic far-field solutions. We connect these solutions to our original potential by scaling. For some special cases, we relate interfacial tensions and line tension to the lengths of the sides and the area of the small triangle formed by the three minima.

\section{Generalized Quartic Potentials} \label{sec:gpqf}

To find more general quartic potentials with arbitrary mimina, while maintaining the condition of constant molar density, we start with a discussion of quartic functions with two independent variables that correspond to two independent mole fractions. In general, a two-variable quartic polynomial can be expressed by
\begin{equation}\label{eq:fxy}
f(X,Y)=\sum\limits_{n=0}^{4} \sum\limits_{i=0}^{n} a_{i,n-i} X^i Y^{n-i},
\end{equation}
which contains 15 independent parameters $a_{i,n-i}$. However, we need a form such that $f$ is positive except for $f=0$ at the three minima, which gives us three equations relating the coefficients. At each of the minima, we require $\partial f/\partial X = \partial f/\partial Y=0$, which amounts to six conditions. Furthermore, in order to have parabolic potential wells, we need the second derivatives of $f$ to satisfy the following inequalities at the three minima: $\partial^2 f/\partial X^2>0$ (or $\partial^2 f/\partial Y^2 > 0$) and $(\partial^2 f/\partial X^2)(\partial^2 f/\partial Y^2)-(\partial^2 f/\partial X\partial Y)^2 >0$. This gives us six inequalities. In general, we have $15-3-6=6$ free parameters together with six additional constraints.

In this framework, our original potential in \eqref{eq:potential0} is a special case which has only one parameter, $a$, representing the size or orientation (magnification or inversion) of the equilateral triangle formed by the three minima \cite[sec.~II.A, discussion after eq.~14]{lin2012mean}. In the development to follow, we obtain a positive potential by assuming it to be a sum of squares of various expressions. We break the symmetry of our original potential by locating the three minima at the vertices of a small triangle having any shape and orientation within the Gibbs triangle. In the following discussion, we use the term a ``small triangle'' to denote an inner triangle formed by the three minima of a given potential within the Gibbs triangle.

\subsection{First Generalization}

At first, we explore the structure of our original potential $f^o$ in \eqref{eq:potential0}. $X_i$ can be expressed as a function of two independent Cartesian variables $u$ and $v$, i.e. $X_i=X_i(u,v)$ as illustrated in Fig.~\ref{subfig:gibbsspace}. In \cite{lin2012mean}, we showed that $f^o$ can be scaled to the form
\begin{equation}\label{eq:finxcorner}
\tilde{f}= \frac{f^o}{(a-b)^4}=\sum\limits_{i=1}^3 Y_i(u,v)^2(Y_i(u,v)-1)^2,
\end{equation}
where $Y_i(u,v)=(X_i(u,v)-b)/(a-b)$ are scaled mole fractions, and $\sum_{i=1}^{3} Y_i=1$. This form is equivalent to $f^o$ when $a=1$. For simplicity, we mainly compare new potentials to $\tilde{f}$.

\begin{figure}[htp]
\centering
\subfigure[\; Geometry of the three minima of $\tilde{f}$ (or $f^o$ when $a=1$)]{
\label{subfig:gibbsminimacornerwithyi}
\includegraphics[scale=0.5]{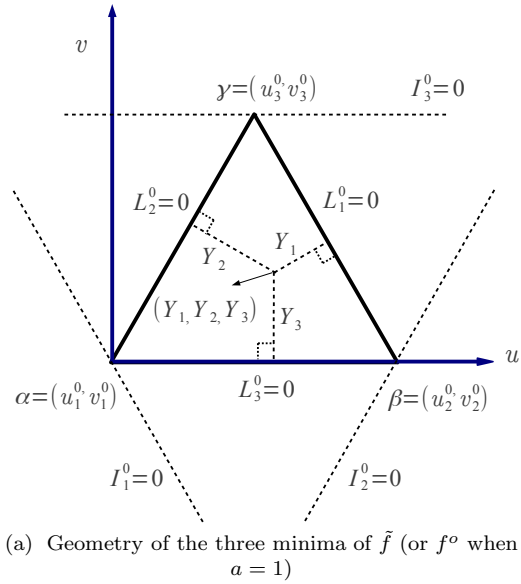}}
\subfigure[\; Contours of $\tilde{f}$]{
\label{subfig:gibbsminimacornersample}
\includegraphics[scale=0.48]{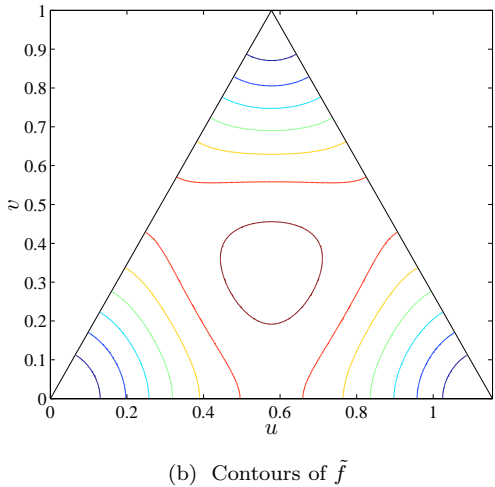}}
\caption{\subref{subfig:gibbsminimacornerwithyi} Geometry of the three minima of the potential $\tilde{f}$ given by \eqref{eq:finxcorner} (or $f^o$ in \eqref{eq:potential0} when $a=1$). $\tilde{f}$ is a function of scaled mole fractions $\{Y_i\}_{i=1,2,3}$, where $\sum_{i=1}^{3} Y_i=1$. $Y_i$ can be expressed in terms of two independent Cartesian coordinates $(u,v)$. $\alpha=(u_1^0,v_1^0)=(0,0)$, $\beta=(u_2^0,v_2^0)=(2/\sqrt{3},0)$, and $\gamma=(u_3^0,v_3^0)=(1/\sqrt{3},1)$ are the three minima of $\tilde{f}$ located at the three corners of the Gibbs triangle. $L_i=0$ represents a line that passes through two points $(u_j^0,v_j^0)$ and $(u_k^0,v_k^0)$ $(j \neq k\neq i)$, and $I_i=0$ is a line parallel to $L_i=0$ but passing through only one point $(u_i^0,v_i^0)$. \subref{subfig:gibbsminimacornersample} Contour plot of $\tilde{f}$. The value of $\tilde{f}$ decreases from its value at the center of the inner area as one proceeds toward the three vertices where it is zero.}
\end{figure}

As illustrated in Fig.~\ref{subfig:gibbsminimacornerwithyi}, we specify explicitly the Cartesian coordinates $(u,v)$ for $Y_i(u,v)$
\begin{equation}\label{eq:xtouv}
\begin{split}
Y_1(u,v) &=-\frac{\sqrt{3}}{2} u - \frac{v}{2} +1 \\
Y_2(u,v) &=\frac{\sqrt{3}}{2} u - \frac{v}{2} \\
Y_3(u,v) &= v.
\end{split}
\end{equation}
In \eqref{eq:xtouv}, the symbols $u$ and $v$ bear the same relationship to the scaled mole fractions $Y_i$ as $u$ and $v$ in Fig~\ref{subfig:gibbsspace} do to the unscaled mole fractions $X_i$. The three minima of the potential $\tilde{f}$ in \eqref{eq:finxcorner}, namely $\alpha=(u_1^0,v_1^0)=(0,0)$, $\beta=(u_2^0,v_2^0)=(2/\sqrt{3},0)$, and $\gamma=(u_3^0,v_3^0)=(1/\sqrt{3},1)$ are located at the three corners of the Gibbs triangle as in Fig.~\ref{subfig:gibbsminimacornerwithyi}, consistent with the contour plot in Fig.~\ref{subfig:gibbsminimacornersample}. Note that $Y_i(u_j^0,v_j^0)=Y_i(u_k^0,v_k^0)=0$ for $j \neq k \neq i$, and $Y_i(u_i^0,v_i^0)=1$.

If we define two sets of linear functions $L_i^0(u,v)\equiv Y_i(u,v)$ and $I_i^0(u,v) \equiv Y_i(u,v)-1 = L_i^0(u,v)-1$, the form \eqref{eq:finxcorner} of the potential $\tilde{f}$ can be expressed as
\begin{equation}\label{eq:finLI}
\tilde{f}=\sum\limits_{i=1}^3 L_i^0(u,v)^2 I_i^0(u,v)^2.
\end{equation}
As shown in Fig.~\ref{subfig:gibbsminimacornerwithyi}, $L_i^0=0$ is a line in the Gibbs space that passes through two minima $(u_j^0,v_j^0)$ and $(u_k^0,v_k^0)$, where $j \neq k \neq i$, and $I_i^0=0$ is a line parallel to $L_i=0$ that passes through the remaining minimum $(u_i^0,v_i^0)$. For each minimum of $\tilde{f}$, there are three intersecting lines.

Based on the geometrical interpretation of the form \eqref{eq:finLI} of the potential $\tilde{f}$, we can generalize its structure by allowing these pairs of parallel lines to move and requiring the intersections of three lines to be located within the Gibbs triangle. To generate a potential function with the desired properties, we locate the three minima arbitrarily,  $\alpha=(u_1,v_1)$, $\beta=(u_2,v_2)$, and $\gamma=(u_3,v_3)$, within the Gibbs triangle the potential . Then, we define two sets of linear functions $L_i(u,v)$ and $I_i(u,v)$, where $L_i(u,v)=0$ is a line passing through two minima $(u_j,v_j)$ and $(u_k,v_k)$ for $j \neq k \neq i$, and $I_i(u,v)=0$ is a line parallel to $L_i(u,v)=0$ and passing through the remaining minimum $(u_i,v_i)$.

\begin{figure}[h!]
\centering
\subfigure[\; Geometry of the three minima of $f_g$]{
\label{subfig:gibbsminimatlt}
\includegraphics[scale=0.32]{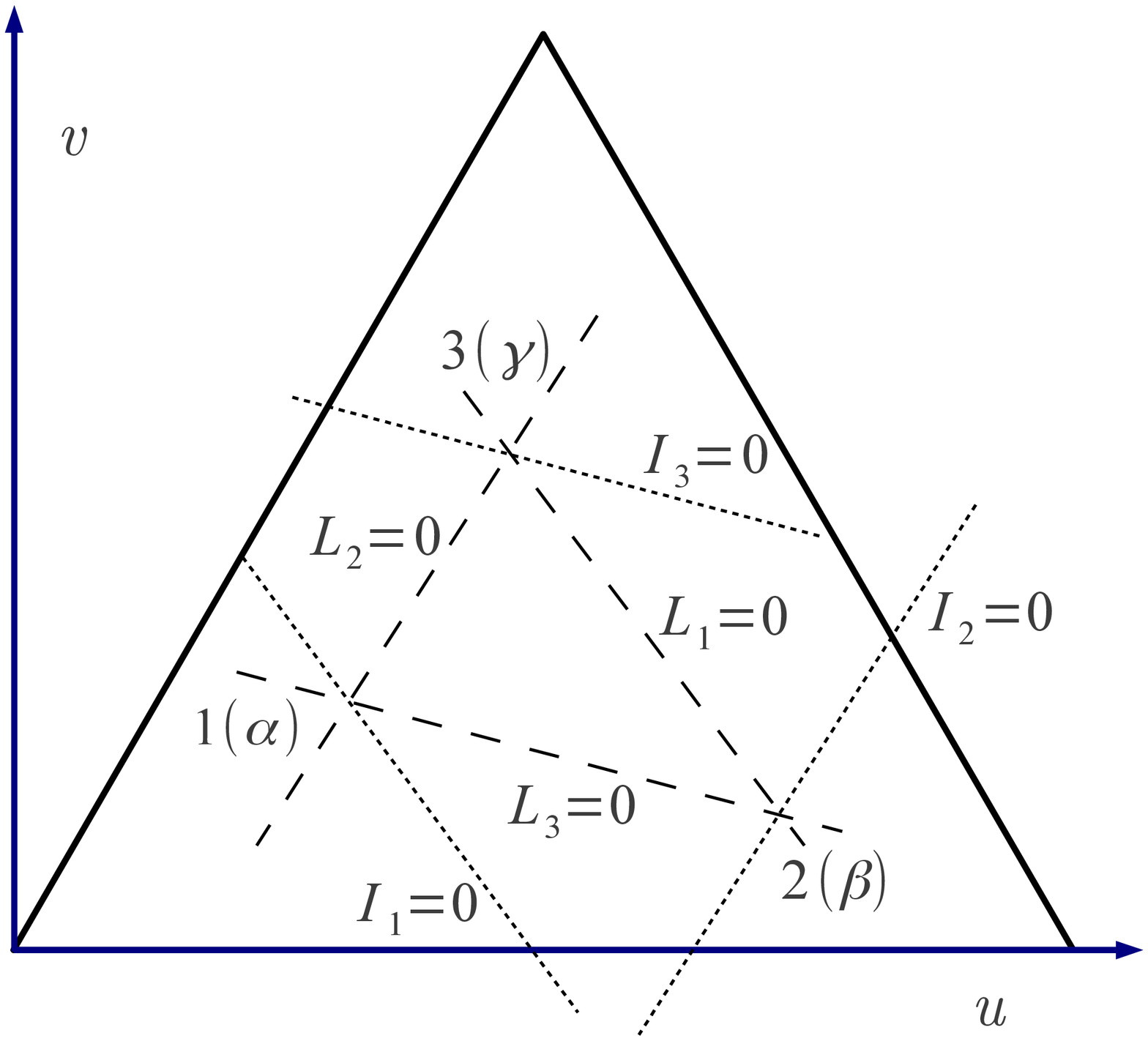}}
\subfigure[\; Contours of $f_g$]{
\label{subfig:gibbsminimatltsample}
\includegraphics[scale=0.48]{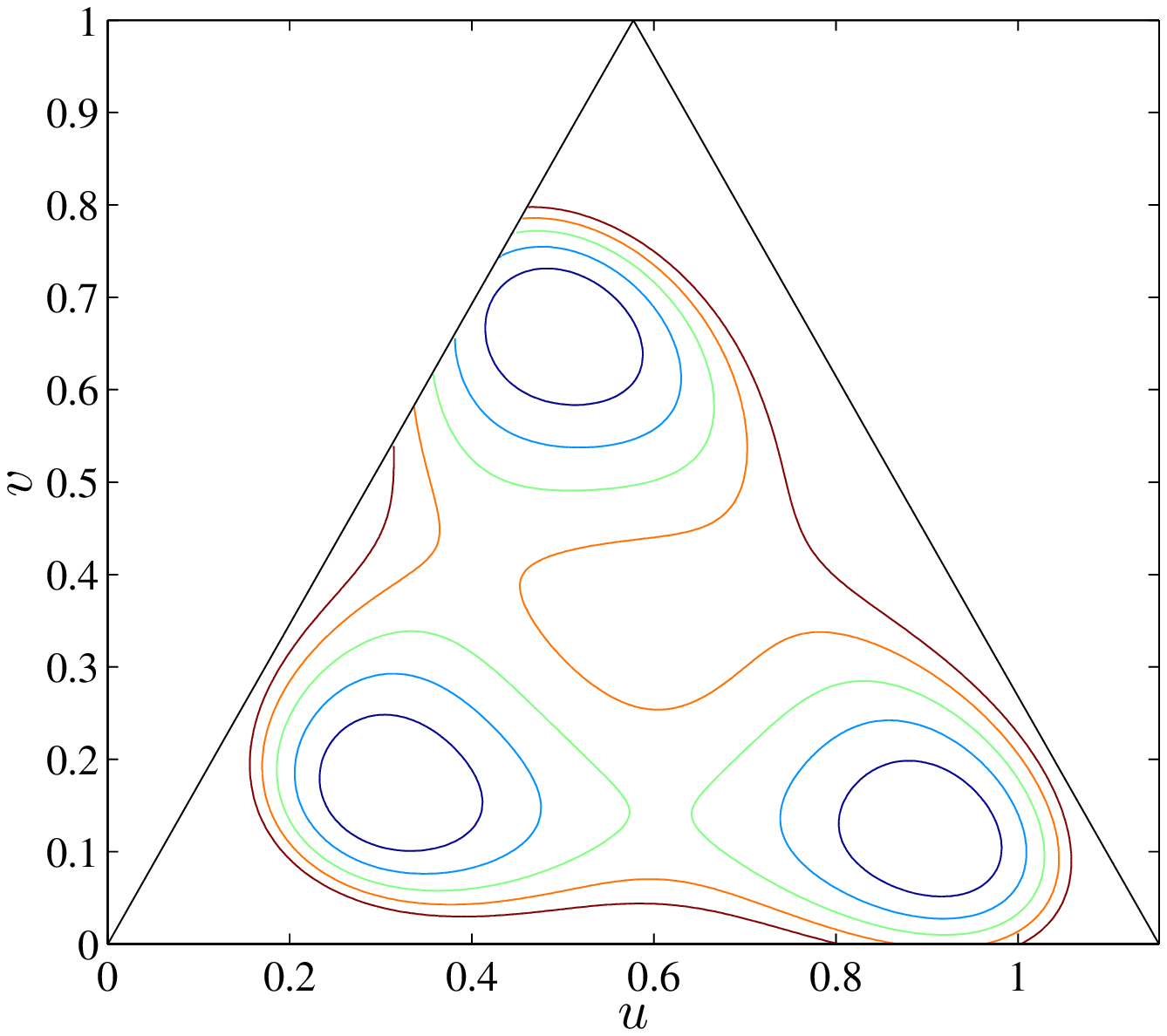}}
\caption{\subref{subfig:gibbsminimatlt} Geometry of the three minima of the quartic potential $f_g$ given by \eqref{eq:finLIgeneralized}. $\alpha=(u_1,v_1)$, $\beta=(u_2,v_2)$, and $\gamma=(u_3,v_3)$ are the three minima of this potential arbitrarily located at any three points within the Gibbs triangle. $L_i=0$ represents a line that passes through $(u_j,v_j)$ and $(u_k,v_k)$ for $j \neq k \neq i$, and $I_i=0$ is a line parallel to $L_i=0$ but passing through the vertex $(u_i,v_i)$ \subref{subfig:gibbsminimatltsample} Contour plot of potential $f_g$ for $(u_1,v_1)=(0.3087,0.1667)$, $(u_2,v_2)=(0.9060,0.1167)$, $(u_3,v_3)=(0.4974,0.6667)$, and $ (d_1,d_2,d_3) =(1.0,1.1,1.2)$. }
\end{figure}

As illustrated in Fig.~\ref{subfig:gibbsminimatlt}, the line $L_i=0$ coincides with the side of the small triangle opposite to the vertex $i$, while the line $I_i=0$ passes through the vertex $i$ and is parallel to the line $L_i=0$. Thus, a generalized potential with arbitrary mimina is given by
\begin{equation}\label{eq:finLIgeneralized}
f_g=\sum\limits_{i=1}^3 d_i^2 L_i(u,v)^2 I_i(u,v)^2,
\end{equation}
where the $d_i$ are nonzero weighting coefficients that relate to the curvatures along the lines that connect each pair of potential wells. For each term, $L_i(u,v)^2 I_i(u,v)^2$ represents a positive quartic function that vanishes along two parallel lines $L_i(u,v)=0$ and $ I_i(u,v)=0$. We need all three terms to produce a positive quartic function that vanishes at the three points $(u_i,v_i)_{i=1,2,3}$ which are the intersections of three lines chosen from the three pairs of parallel lines. Contours for such a potential are illustrated in Fig.~\ref{subfig:gibbsminimatltsample}.

The choice of $L_i(u,v)$ and $I_i(u,v)$ for the potential $f_g$ in \eqref{eq:finLIgeneralized} is somewhat arbitrary but we use specific forms of $L_i(u,v)$ and $I_i(u,v)$ and a weighting coefficient for each term to maintain generality. The specific forms of $L_i(u,v)$ and $I_i(u,v)$ that we use are
\begin{equation}\label{eq:LiIi}
\begin{split}
L_i(u,v)  & \equiv C_{ijk}[(u-u_j)v_{jk} - (v-v_j)u_{jk}] \\
& = C_{ijk}[(u-u_k)v_{jk} - (v-v_k)u_{jk}] \\
I_i(u,v) & \equiv L_i(u,v) -h \\
& = C_{ijk} [(u-u_i)v_{jk} - (v-v_i)u_{jk}] ,
\end{split}
\end{equation}
where $C_{ijk} \equiv (\epsilon_{ijk}+\lvert \epsilon_{ijk} \rvert)/2$, $\epsilon_{ijk}$ is the Levi-Civita symbol for three dimensions, $u_{jk} \equiv u_j-u_k$  and $v_{jk} \equiv v_j-v_k$; moreover, $h= L_i(u_i,v_i) = 2 A$, and $A$ is the area of the small triangle. For simplicity, we require the indices of the three vertices to be numbered counterclockwise. Note that
\begin{equation}\label{eq:h}
h = \sum\limits_{(i,j,k)\in \lbrace 1,2,3\rbrace} \epsilon_{ijk}u_jv_k 
= \left\lvert \begin{array}{ccc} 1 & 1 & 1 \\ u_1 & u_2 & u_3 \\ v_1 & v_2 & v_3 \end{array} \right\rvert.
\end{equation}

If we express $L_i$ and $h$ in terms of two independent mole fractions $Y_1$ and $Y_2$, we obtain
\begin{equation}
\begin{split}
L_i & = \frac{2}{\sqrt{3}}C_{ijk}\left[(Y_1-Y_1^{(k)})Y_2^{(jk)} - (Y_2-Y_2^{(k)})Y_1^{(jk)} \right] \\
h & = \frac{2}{\sqrt{3}}\sum\limits_{(i,j,k)\in \lbrace 1,2,3\rbrace} \epsilon_{ijk}Y_1^{(j)}Y_2^{(k)},
\end{split}
\end{equation}
in which we have defined $Y_i^{(jk)} \equiv Y_i^{(j)} - Y_i^{(k)}$ and expressed the three minima $(u_i,v_i)_{i=1,2,3}$ in terms of $(Y_1^{(i)},Y_2^{(i)})_{i=1,2,3}$ by using the relation \eqref{eq:xtouv} of $(u,v)$ and $Y_i$.

The form \eqref{eq:finLIgeneralized} of the potential $f_g$ has three coefficients $d_i$ and six parameters $(u_i,v_i)$ that locate the positions of the three minima, nine parameters in total. One could take out a common factor without changing the nature of the potential. From the discussion in Page~\pageref{eq:fxy} for the desired properties of a general quartic potential given by \eqref{eq:fxy}, we know that there are six free parameters, which means there are three parameters that might be able to be merged with others.

From the forms \eqref{eq:finxcorner} and \eqref{eq:finLI} of the potential $\tilde{f}$, our original potential $f^o$ divided by a factor $(a-b)^4$ can be viewed as a special case of the potential $f_g$ given by \eqref{eq:finLIgeneralized}, in which the three minima form an equilateral small triangle with one parameter $a$ related to its size. With this parameter, we can scale the small triangle from the Gibbs triangle to its center point with zero area. To add degrees of freedom to allow this equilateral small triangle to move around within the Gibbs triangle, we can add two extra parameters corresponding to translations in two perpendicular directions. For arbitrary shape and orientation of the small triangle, we need a parameter corresponding to rotation and two parameters for distortion, such as changes of two inner angles. In general, these operations correspond to a linear transformation with six free parameters, which will be discussed in Section~\ref{sec:lineartransform}.

The general properties of the potential $f_g$ in \eqref{eq:finLIgeneralized} have been tested by performing calculations of its first and second derivatives with respect to $u$ and $v$. For details, see \cite[sec.~5.1.1]{lin2012meanthesis}. From these calculations, we prove that $f_g$, which is a generalization of potential $\tilde{f}$ in \eqref{eq:finxcorner}, satisfies our desired properties of a quartic polynomial, as discussed in the beginning of Sec.~\ref{sec:gpqf}. We also calculated the eigenvalues of the Hessian matrix $H$, which is the matrix in terms of the second derivatives of $f_g$ with respect to $u$ and $v$. These eigenvalues $\lambda_{\pm}$ are the principle curvatures at the mimina, i.e.
\begin{equation}\label{eq:eigenvalueshessianfg}
\begin{split}
\lambda_{\pm}= &  h^2 \left\lbrace \sum\limits_{i=1}^{3} d_i^2 S_i^2 \right. \\
\pm & \left. \sqrt{ \left(\sum\limits_{i=1}^{3} d_i^2 S_i^2 \right)^2 - 4 h^2 \sum\limits_{(i,j,k) \in \{ 1,2,3\}} C_{i,j,k} d_j^2 d_k^2 } \right\rbrace,
\end{split}
\end{equation}
where $S_i\equiv C_{ijk} \sqrt{u_{jk}^2 + v_{jk}^2}$ is the length of the side of the small triangle opposite to the vertex $(u_i,v_i)$, as in Fig.~\ref{fig:gibbsminimasigmageometry}. Note that both of its eigenvalues $\lambda_{\pm}$ are positive.

For the case that all $d_i=1$ and $S_1=S_2=S_3=(2/\sqrt{3})(a-b)$, the eigenvalues of $H$ in \eqref{eq:eigenvalueshessianfg} reduce to
\begin{equation}
\lambda_{\pm}  = \frac{2}{3} (a-b)^6 (2 \pm \sqrt{3})
\end{equation}
which connects to our original potential $f^o$ in \eqref{eq:potential0}. When $a=b=1/3$, these two eigenvalues vanish, which means the curvatures vanish. This is equivalent to the bulk criticality for $f^o$, for which the three minima merge to a one minimum located at the center of the Gibbs triangle. In this case, the three mole fractions are uniform throughout the physical space, i.e. $X_i=1/3$. The resulting intefacial and line tensions at the vicinity of this critical value scale consistently with previously determined results in our earlier work \cite{lin2012mean}, in which the ratio of their exponents in terms of $\lvert X_i-1/3 \rvert$ satisfies the mean-field approximation in \cite{varea1992statistical}.

\subsection{Second Generalization}

\begin{figure}[h!]
\centering
\subfigure[\; Geometry of the three minima of $f_G$]{
\label{subfig:gibbsminimageneral}
\includegraphics[scale=0.32]{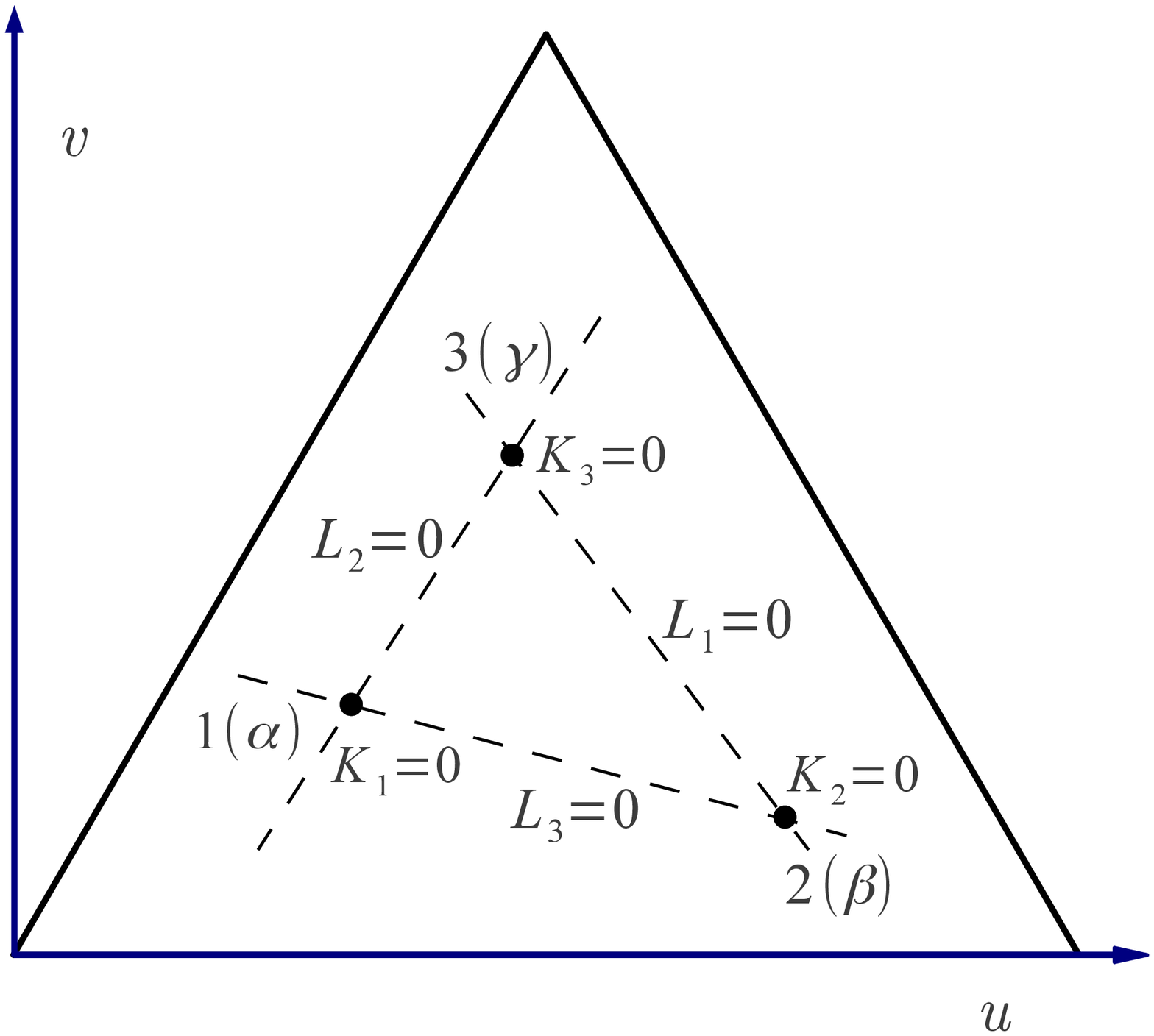}}
\subfigure[\; Contours of $f_G$]{
\label{subfig:gibbsminimageneralsample}
\includegraphics[scale=0.5]{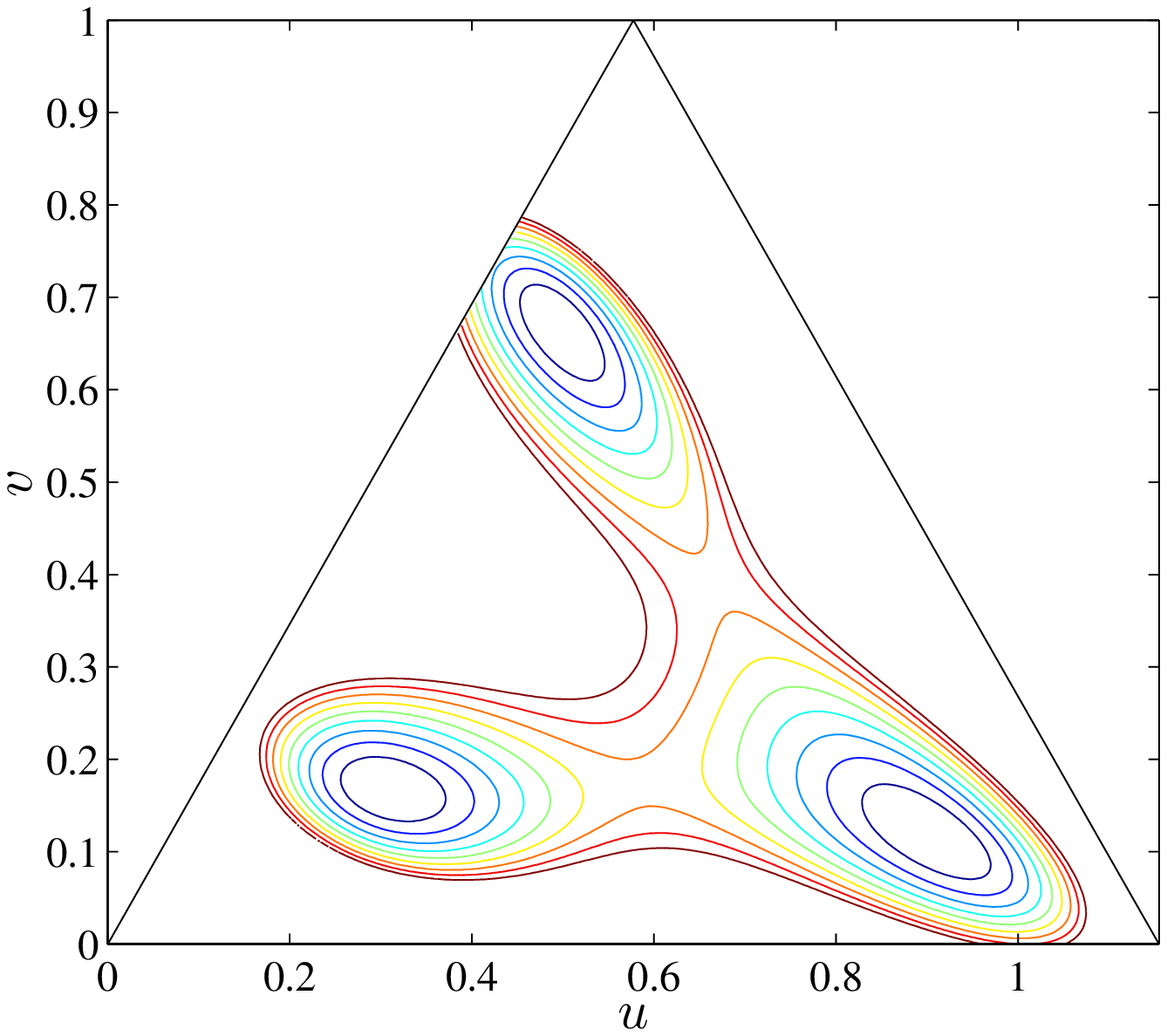}}
\caption{\subref{subfig:gibbsminimageneral} Geometry of the three minima of the quartic potential $f_G$ given by \eqref{eq:finLK}. $\alpha=(u_1,v_1)$, $\beta=(u_2,v_2)$, and $\gamma=(u_3,v_3)$ are the three minima of this potential arbitrarily located at any three points within the Gibbs triangle. $L_i=0$ represents a line passing through $(u_j,v_j)$ and $(u_k,v_k)$, and $K_i=0$ is a point located at $(u_i,v_i)$. \subref{subfig:gibbsminimageneralsample} Contour plot of potential $f_G$ for $(u_1,v_1)=(0.3087,0.1667)$, $(u_2,v_2)=(0.9060,0.1167)$, $(u_3,v_3)=(0.4974,0.6667)$, and $ (a_i,b_i,c_i)=(1,1,1)$, $(i=1,2,3)$.}
\end{figure}

The potential $f_g$ in \eqref{eq:finLIgeneralized} can be generalized by replacing $I_i$ with an arbitrary linear function that passes through a point $(u_i,v_i)$, which is not necessarily parallel ot $L_i$. For details, see Appendix~\ref{sec:appendixA}. We also observed that $I_i(u,v)^2$ of $f_g$ does not have to vanish on a line $I_i(u,v)=0$. It can be replaced by a positive quadratic function $K_i(u,v)$ which is a paraboloid in three dimensional space that vanishes only at the point $(u_i,v_i)$. In general, $K_i(u,v)$ can be expressed as
\begin{equation}\label{eq:ki}
K_i (u,v) \equiv [a_i(u-u_i) + b_i(v-v_i)]^2 + c_i^2 (v-v_i)^2,
\end{equation}
where $a_i$, $b_i$, and $c_i$ are three coefficients, and $c_i \ne 0$. Then a more general potential, as illustrated in Fig.~\ref{subfig:gibbsminimageneral}, can be expressed in the form
\begin{equation}\label{eq:finLK}
f_G=\sum\limits_{i=1}^3 L_i(u,v)^2 K_i(u,v).
\end{equation}

$f_G$ in \eqref{eq:finLK} is the most general potential we have explored. It has nine coefficients and six parameters $(u_i,v_i)$ for the positions of the three minima. Thus, it has 15 coefficients in total. We note that when $a_i = C_{ijk} d_i v_{jk}$, $b_i = -C_{ijk} d_i u_{jk}$, and $c_i=0$, $f_G$ reduces to the potential $f_g$ in \eqref{eq:finLIgeneralized}. Therefore, $f_g$ is a subset of $f_G$, where $f_G$ has six extra parameters corresponding to the orientations of the three potential wells, as well as the general shape of the potential. Recall the properties for the desired potential: the potential has to be positive throughout the entire domain and has three minima at zero potential; at these minima, the first derivatives vanish and the Hessien matrices for second derivatives are positive definite. These properties have been tested in \cite[sec.~5.1.2]{lin2012meanthesis}.

\begin{figure}[h!]
\centering
\includegraphics[scale=0.44]{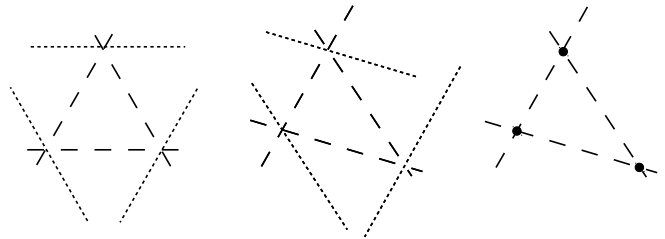}
\caption{Schematic representation of the generalization from a potential $\tilde{f}$ in \eqref{eq:finxcorner} to the potential $f_g$ in \eqref{eq:finLIgeneralized} of similar form and to the more general potential $f_G$ in \eqref{eq:finLK}. At the LHS of this figure, there are three pairs of two parallel lines. The three minima of $\tilde{f}$ are located at the three nodes where lines intersect and form a small equilateral triangle. In the middle of this figure, we generalize $\tilde{f}$ to $f_g$ by allowing the three minima of $f_g$ to be at arbitrary locations. In this case, the three pairs of parallel lines determine the three minima. The RHS of this figure shows a further generalization in which each minimum of $f_G$ is located by two lines and a point.
}
\label{fig:gibbsminimaevolution}
\end{figure}

Fig.~\ref{fig:gibbsminimaevolution} presents a schematic summary of the generalization from the quartic potential $\tilde{f}$ in \eqref{eq:finxcorner} (equivalent to our original potential $f^o$ in $\eqref{eq:potential0}$ when $a=1$) to a similar form \eqref{eq:finLIgeneralized} of the potential $f_g$ and to a more general quartic form \eqref{eq:finLK} of the potential $f_G$. In the LHS of Fig.~\ref{fig:gibbsminimaevolution}, we see that the three minima of $\tilde{f}$ form an equilateral small triangle and each minimum is located at an intersection of three lines. These lines consist of three pairs of two parallel lines. For each pair, there is a line passing through two minima and a parallel line passing through the remaining minimum. Each line corresponds to a paraboloid that vanishes at this line, and each pair of parallel lines corresponds to the minima of a quartic function. Thus, $\tilde{f}$ is a sum of three quartic function. As a generalization, in the middle of Fig.~\ref{fig:gibbsminimaevolution}, the three minima of $f_g$ form a small triangle of arbitrary shape. The rules of selecting parallel pairs of lines and constructing corresponding quartic functions for $f_g$ are the same as for $\tilde{f}$. The RHS of Fig.~\ref{fig:gibbsminimaevolution} shows a further generalized potential $f_G$, where each minimum is not located at an intersection of three lines but two lines and a point. In this case, each pair of parallel lines in $f_g$ is transformed to a line passing through two minima and a point at the remaining minimum. To compose $f_G$, we need two sets of quadratic functions. The one in first set vanishes along a line and the one in the second set vanishes at a point.

\section{Generalized Potentials with Analytic Far-field Solutions}

\subsection{Straight line trajectory of two-phase transition}\label{sec:straightlinetrajectory}

In section~\ref{sec:gpqf}, we introduced a systematic way to find quartic generalized potentials by means of geometric considerations. In order to find the asymptotic solutions for two-phase transitions in physical space for each of the three interfaces in the regions far from the three-phase contact line, we need to know the corresponding trajectories for transitions between the pairs of wells of a given potential within the Gibbs triangle. According to the discussion in \cite[sec.~2.4.1]{lin2012meanthesis}, the property of our original potential that leads us to simple far-field analytic solutions is that the trajectory of a transition between any two phases follows a straight line within the Gibbs triangle. From a geometric aspect, this straight line lies along a valley connecting a pair of minima; as one goes along this valley, the potential rises from a minimum, goes over a saddle point, and decreases, to another minimum. The first derivative and the second derivative of the potential with respect to the normal of this straight line in the Gibbs space are zero and positive, respectively. For the subset of generalized potentials that share this common property, we can find analytical far-field solutions. As shown in Appendix~\ref{sec:appendixBformwithanalyticsoln}, the form of this subset is actually equivalent to the form \eqref{eq:finLIgeneralized} of the generalized potential $f_g$.

\subsection{Linear transformation}\label{sec:lineartransform}

According to the geometric representation in section~\ref{sec:gpqf}, as illustrated in Fig.~\ref{fig:gibbsminimaevolution}, the form \eqref{eq:finLI} of the potential $\tilde{f}$ can be generalized to a similar quartic potential $f_g$ in \eqref{eq:finLIgeneralized} with arbitrary minima $(u_1,v_1)$, $(u_2,v_2)$, and $(u_3,v_3)$ by choosing proper linear functions $L_i(u,v)$ and $I_i(u,v)$. In this way, the obtained potential has six parameters $\lbrace u_i,v_i \rbrace_{i=1,2,3}$ and three weighting coefficients $\lbrace d_i \rbrace_{i=1,2,3}$, which match our earlier reasoning about using six parameters that correspond to operations such as scaling, translation, rotation, and distortion. Furthermore, we notice that these operations may be represented by linear transformations of our original potential.

\begin{figure}[h!]
\centering
\includegraphics[scale=0.4]{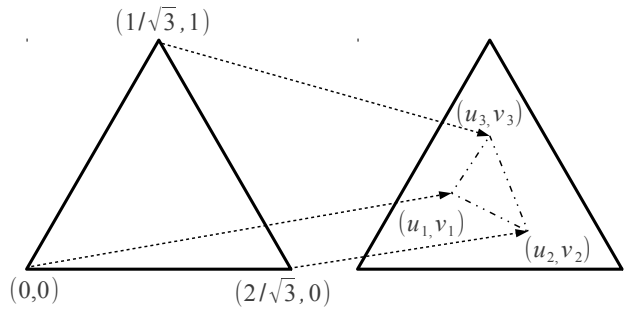}
\caption{Schematic linear transformation from the three corners of the Gibbs triangle to three arbitrary internal points $(u_1,v_1)$, $(u_2,v_2)$, and $(u_3,v_3)$ within it.}
\label{fig:gibbsminimalineartransf}
\end{figure}

To construct a linear transformation that maps the three minima from the corners of the Gibbs triangle to three arbitrary internal points $(u_1,v_1)$, $(u_2,v_2)$, and $(u_3,v_3)$ as illustrated in Fig.~\ref{fig:gibbsminimalineartransf}, we write 
\begin{equation}
\begin{bmatrix}
u' \\ v'
\end{bmatrix}
= \begin{bmatrix}
\alpha & \beta \\ \gamma & \delta
\end{bmatrix}
\begin{bmatrix}
u \\ v
\end{bmatrix}
+
\begin{bmatrix}
\xi \\ \eta
\end{bmatrix},
\end{equation}
where $\alpha$, $\beta$, $\gamma$, $\delta$, $\xi$, and $\eta$ are six undetermined parameters. By substituting $(u_1,v_1)$, $(u_2,v_2)$, and $(u_3,v_3)$ for $(u',v')$, and the corresponding three corners $(0,0)$, $(2/\sqrt{3},0)$, and $(1/\sqrt{3},1)$ for $(u,v)$, we solve
\begin{equation}
\begin{bmatrix}
u' \\ v'
\end{bmatrix}
= \begin{bmatrix}
-\frac{\sqrt{3}}{2}u_{12}& \frac{1}{2}u_{12} +u_{31} \\ -\frac{\sqrt{3}}{2}v_{12}& \frac{1}{2}v_{12} + v_{31}
\end{bmatrix}
\begin{bmatrix}
u \\ v
\end{bmatrix}
+
\begin{bmatrix}
u_1 \\ v_1
\end{bmatrix},
\end{equation}
where $u_{ij}\equiv u_i- u_j$ and $ v_{ij}\equiv v_i- v_j$. Then, by matrix inversion, we find
\begin{equation}\label{eq:uvtouvprime}
\begin{bmatrix}
u \\ v
\end{bmatrix}
= \frac{2}{\sqrt{3} h}\begin{bmatrix}
\frac{1}{2}v_{12} + v_{31} & - \frac{1}{2}u_{12} - u_{31} \\ \frac{\sqrt{3}}{2}v_{12}& -\frac{\sqrt{3}}{2}u_{12}
\end{bmatrix}
\begin{bmatrix}
u'-u_1 \\ v'-v_1
\end{bmatrix},
\end{equation}
where $h \neq 0$ is twice the area of the small triangle, defined in \eqref{eq:LiIi} and \eqref{eq:h}.

A generalization of the potential $\tilde{f}$ in \eqref{eq:finxcorner} with weighting coefficients $e_i$ can be expressed in terms of two independent Cartesian mole fractions $u$ and $v$ as follows:
\begin{equation}\label{eq:fdaggerinuv} 
\begin{split}
f^{\dagger}(u,v) & \equiv f^{\dagger}(Y_1(u,v), Y_2(u,v),Y_3(u,v)) \\
& \equiv \sum\limits_{i=1}^{3} e_i^2 Y_i(u,v)^2 (Y_i(u,v)-1)^2.
\end{split}
\end{equation}
To construct a new potential with three minima located at $(u_1,v_1)$, $(u_2,v_2)$, and $(u_3,v_3)$, we replace $Y_i(u,v)$ in \eqref{eq:fdaggerinuv}  by 
\begin{equation}
Z_i(u',v')\equiv Y_i(u(u',v'),v(u',v')),
\end{equation}
where the linear transformation \eqref{eq:uvtouvprime} from $(u',v')$ to $(u,v)$ has been used. Then (see \cite{lin2012meanthesis} for details), we find (after dropping the primes)
\begin{equation}\label{eq:zi}
Z_i(u,v) = \frac{L_i(u,v)}{h},
\end{equation}
where, as defined in \eqref{eq:LiIi}, $L_i$ is a linear function that satisfies $L_i(u_j,v_j)=0$ and $L_i(u_k,v_k)=0$ for $(j \neq k \neq i)$; whereas $L_i(u_i,v_i)=h$, which leads to $Z_i(u_i,v_i)=1$. Note that $\sum_{i=1}^3 Z_i =1$. 
Thus, the new potential is defined as
\begin{equation}\label{eq:fast}
\begin{split}
f^{\ast}(u,v) & \equiv f^{\dagger}(Z_1(u,v),Z_2(u,v),Z_3(u,v)) \\
& = \sum_{i=1}^3 e_i^2 Z_i(u,v)^2(Z_i(u,v)-1)^.
\end{split}
\end{equation}

With the definition \eqref{eq:fast} of the new potential $f^{\ast}$, we find that $\{f^{\ast}(u_i,v_i)=0\}_{i=1,2,3}$, which shows that the three mimina of $f^{\ast}$ are located at $\lbrace(u_i,v_i)\rbrace_{i=1,2,3}$. By expressing $f^{\ast}$, which relates to the potential $f^{\dagger}$ in \eqref{eq:fdaggerinuv} (generalized from the potential $\tilde{f}$ in \eqref{eq:finxcorner}) by a linear transformation, in terms of $L_i$, i.e.
\begin{equation}
f^{\ast} = \sum_{i=1}^3 \frac{e_i}{h^4}^2 L_i^2(L_i-h)^2 = \sum_{i=1}^3 d_i^2 L_i^2I_i^2 = f_g,
\end{equation}
where $d_i \equiv e_i/h^2$, we obtain exactly the potential $f_g$ in \eqref{eq:finLI}. Thus, $f_g$ is a linear transformation from the original potential $\tilde{f}$ with extra weighting coefficients. This is the reason that $\tilde{f}$ and $f_g$ share a common property: the two-phase transitions in the physical domain follow straight lines in the Gibbs space.

Notice that the structure of the two-density quartic potential used by Widom \textit{et al} \cite{szleifer1992surface,koga2010first} actually belongs to the general category indicated by the form \eqref{eq:finLIgeneralized} of the potential $f_g$. The two relative densities of their potential can be treated as our two Cartesian mole fractions $u$ and $v$. We can scale their potential in terms of three dimensionless quantities similar to the $Z_i$ in our expression and obtain a form similar to $f_g$. However, their systems do not follow the assumption of uniform molar volume. Therefore, in their systems, the sum of the quantities $\sum_{i=1}^3 Z_i \neq 1$ and cannot be mapped into our potentials.

\subsection{Scaling}\label{sec:scaling}

From Sec.~\ref{sec:lineartransform}, we know that the potential $f_g$ in \eqref{eq:finLIgeneralized} and the potential $\tilde{f}$ in \eqref{eq:finxcorner} can be connected by a linear transformation. In general, a problem involved $f_g$ can be solved by scaling into a form like $\tilde{f}$, which we studied in \cite[sec.~IV.C]{lin2012mean}. As an example, we consider the excess grand potential given by \eqref{eq:omegaxsgeneral} but with $f \rightarrow f_g$, namely
\begin{equation}\label{eq:omegaxsuv}
\hat{\Omega}_{xs} = B L \int_A \left[ f_g(u,v) +  g( \nabla u, \nabla v) \right] \mathrm{d}A.
\end{equation}
We express the gradient energy in terms of two Cartesian mole fractions $u$ and $v$, 
\begin{equation}\label{eq:ginuv}
g(\nabla u, \nabla v)=\epsilon_u \lvert \nabla u \rvert^2+ \epsilon_v \lvert \nabla v \rvert^2 + \epsilon_{uv} (\nabla u \cdot \nabla v),
\end{equation}
where $\epsilon_u= (3/8)(\ell_1^2 + \ell_2^2)$, $\epsilon_v= (1/8)(\ell_1^2 + \ell_2^2) + (1/2)\ell_3^2$, and $\epsilon_{uv}=(\sqrt{3}/4)(\ell_1^2 - \ell_2^2)$. For isotropic gradient energy, $\ell_i \equiv \ell$, we have $\epsilon_u = \epsilon_v = (3/4)\ell^2$ and $\epsilon_{uv}=0$. According to the scaling \eqref{eq:zi} from $(u,v)$ to $L_i$, we can express $\hat{\Omega}_{xs}$ as
\begin{equation}\label{eq:omegaxsuvh}
\begin{split}
\hat{\Omega}_{xs} & = B L  \int_A \left[\sum_{i=1}^3  d_i^2 L_i^2 I_i^2 + g(\nabla u, \nabla v) \right] \mathrm{d}A \\
& = B L h^4 \int_A \sum_{i=1}^3 \left[ d_i^2 Z_i^2(Z_i-1)^2 +  \frac{\hat{\ell}_i^2}{2} \lvert \nabla Z_i\rvert^2 \right] \mathrm{d}A,
\end{split}
\end{equation}
where
\begin{equation}\label{eq:ellinuv}
\begin{split}
\hat{\ell}_1^2 & = - \frac{2}{h^4}[\epsilon_u u_{12} u_{31} + \epsilon_v v_{12} v_{31} + \frac{\epsilon_{uv}}{2} (u_{12} v_{31} + v_{12} u_{31} )]\\
\hat{\ell}_2^2 & = - \frac{2}{h^4}[\epsilon_u u_{23} u_{12} + \epsilon_v v_{23} v_{12} + \frac{\epsilon_{uv}}{2} (u_{23} v_{12} + v_{23} u_{12} )]\\
\hat{\ell}_3^2 & = - \frac{2}{h^4}[\epsilon_u u_{31} u_{23} + \epsilon_v v_{31} v_{23} + \frac{\epsilon_{uv}}{2} (u_{31} v_{23} + v_{31} u_{23} )].
\end{split}
\end{equation}

\subsubsection{Case: $d_i=1$}

Consider a special case of the potential $f_g$ in \eqref{eq:finLIgeneralized} when all of the weighting coefficients are equal to one, i.e. $d_i=1$. According to the asymptotic analysis in our previous work \cite[sec.~II.C]{lin2012mean}, we know how to obtain analytical solutions for two-phase transitions in a region far from the three-phase contact line for the form \eqref{eq:omegaxsuvh} of the excess grand potential $\hat{\Omega}_{xs}$. These solutions show that the interfacial tension $\hat{\sigma}_{\alpha \beta}$ is proportional to $\sqrt{\hat{\ell}_1^2 + \hat{\ell}_2^2}$, and the similar relations applied to $\hat{\sigma}_{\gamma \alpha}$ and $\hat{\sigma}_{\beta \gamma}$ (see \cite[eq.~25]{lin2012mean}). So we can express the relation \eqref{eq:eqangle} of the dihedral angles and the interfacial tensions as following:
\begin{equation}\label{eq:sinthetaintildeell}
\frac{\sin \theta_{\alpha}}{\sqrt{\hat{\ell}_2^2 + \hat{\ell}_3^2}} = \frac{\sin \theta_{\beta}}{\sqrt{\hat{\ell}_3^2 + \hat{\ell}_1^2}} = \frac{\sin \theta_{\gamma}}{\sqrt{\hat{\ell}_1^2 + \hat{\ell}_2^2}},
\end{equation}
in which the denominators can be obtained by calculating the sums of pairs of $\hat{\ell}_i^2$ in \eqref{eq:ellinuv},
\begin{equation}\label{eq:twotildeellsq}
\begin{split}
\hat{\ell}_1^2 +\hat{\ell}_2^2& = \frac{2}{h^4}[\epsilon_u u_{12}^2 + \epsilon_v v_{12}^2 + \epsilon_{uv} u_{12} v_{12}]\\
\hat{\ell}_2^2 +\hat{\ell}_3^2& = \frac{2}{h^4}[\epsilon_u u_{23}^2 + \epsilon_v v_{23}^2 + \epsilon_{uv} u_{23} v_{23}]\\
\hat{\ell}_3^2 +\hat{\ell}_1^2& = \frac{2}{h^4}[\epsilon_u u_{31}^2 + \epsilon_v v_{31}^2 + \epsilon_{uv} u_{31} v_{31}].
\end{split}
\end{equation}

\subsubsection{Case: $d_i=1$ and $\ell_i=\ell$}\label{sec:dieq1ellieqell}

Following by the discussion in Sec.~\ref{sec:dieq1ellieqell} for the special case with equal weighting coefficients $d_i=1$ of the potential $f_g$ in \eqref{eq:finLIgeneralized}, a subset of this special case posses an interesting property. This subset is specified by assuming the gradient energy $g$ in \eqref{eq:ginuv} is isotropic, given by $\ell_i=\ell$. Then, the sums of pairs of $\hat{\ell}_i^2$ in \eqref{eq:twotildeellsq} become
\begin{equation}\label{eq:twotildeellsqsimple}
\begin{split}
\hat{\ell}_1^2 +\hat{\ell}_2^2& = \frac{3\ell^2}{2h^4}[u_{12}^2 + v_{12}^2]=\frac{3\ell^2}{2h^4}S_3^2\\
\hat{\ell}_2^2 +\hat{\ell}_3^2& = \frac{3\ell^2}{2h^4}[u_{23}^2 + v_{23}^2]=\frac{3\ell^2}{2h^4}S_1^2\\
\hat{\ell}_3^2 +\hat{\ell}_1^2& = \frac{3\ell^2}{2h^4}[u_{31}^2 + v_{31}^2]=\frac{3\ell^2}{2h^4}S_2^2,
\end{split}
\end{equation}
where $S_i$ is the length of the side of the small triangle opposite to the vertex $(u_i,v_i)$, defined earlier in the expression \eqref{eq:eigenvalueshessianfg} for the eigenvalues $\lambda_{\pm}$ of the Hessian matrix of $f_g$. Then, the relation of dihedral angles in \eqref{eq:sinthetaintildeell} reduces to
\begin{equation}\label{eq:sinthetainsi}
\frac{\sin \theta_{\alpha}}{S_1} = \frac{\sin \theta_{\beta}}{S_2} = \frac{\sin \theta_{\gamma}}{S_3}.
\end{equation}
Therefore, for this special case, the sine of each dihedral angle of a bulk phase is proportional to the length of the side of the small triangle opposite to the vertex corresponding to the given bulk phase. This is illustrated in Fig.~\ref{fig:gibbsminimasigmageometry} and will be revisited in Sec.~\ref{sec:caseequalweightisotropic}.

Notice that, in general, $\hat{\ell}_1 \neq \hat{\ell}_2 \neq \hat{\ell}_3$ in \eqref{eq:twotildeellsqsimple}, which leads to the lengths $S_i$ of the three edges of the small triangle unequal. According to the relation \eqref{eq:sinthetainsi} of the dihedral angles and $S_i$, the three dihedral angles are different. Therefore, $f_g$ is capable of representing an asymmetric three-phase contact line with isotropic gradient energy

\subsection{Asymptotic far-field solutions}\label{sec:fgasymptoticsolution}

Alternatively, we can find the relationship of the dihedral angles of a system specified by the potential $f_g$ in \eqref{eq:finLIgeneralized} directly. We consider a transition from phase $\beta=(u_2,v_2)$ to phase $\alpha=(u_1,v_1)$, which follows a straight line $L_3=0$ within the Gibbs triangle as illustrated in Fig.~\ref{subfig:gibbsminimatlt}. In the far-field limit, the transition occurs in a regime far from the three-phase contact line in the physical space. In this regime, the three mole fractions $Y_i$ satisfy the boundary condition $\nabla Y_i \cdot \hat{n}=0$. By assuming $u_{12} \neq 0$, we can interchange $v$ with $u$ according to $L_3=0$, i.e. $v-v_2=(v_{12}/u_{12})(u-u_2)$ and $v-v_1=(v_{12}/u_{12})(u-u_1)$. Then the problem is essentially a one-dimensional problem. The excess grand potential $\hat{\Omega}_{xs}$ in \eqref{eq:omegaxsuvh} reduces to the form
\begin{equation}\label{eq:omegaxs1dL12}
\hat{\Omega}_{xs} = BLw \left(\frac{h}{u_{12}}\right)^4 \mathlarger{\int} \left[ H(u) + \hat{\epsilon}_{uv} \left( \frac{\mathrm{d}u}{\mathrm{d}s} \right)^2 \right] \mathrm{d}s,
\end{equation}
where $s$ is a coordinate perpendicular to the $\alpha \beta$-interface measured from $\beta$ to $\alpha$ and $w$ is the width of an area in the far field regime as illustrated in the bottom of Fig.~\ref{fig:realgibbscompare} (for details, see \cite[fig.~4]{lin2012mean}), $H(u)\equiv (d_1^2 + d_2^2)(u-u_1)^2(u-u_2)^2$, and $\hat{\epsilon}_{uv}=(u_{12}^2/h^4)(\epsilon_u u_{12}^2 + \epsilon_v v_{12}^2 + \epsilon_{uv} u_{12} v_{12})$. The limits of integration are effectively from $-\infty$ to $\infty$.

\begin{figure}[h!]
\centering
\includegraphics[scale=0.54]{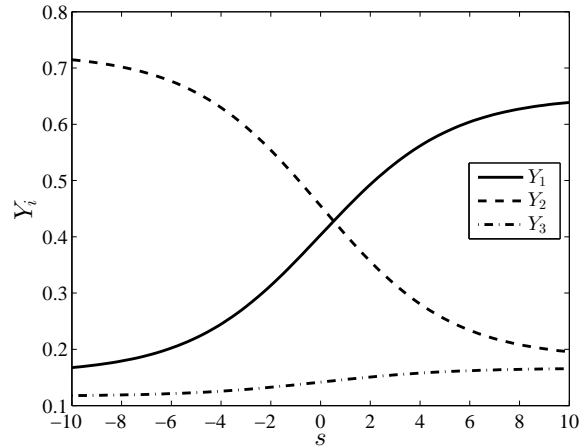}
\caption{Analytic far-field solutions for the mole fraction $Y_i$ for a special case of the potential $f_g$ in \eqref{eq:finLIgeneralized} at the $\alpha \beta$-interface. In this special case, $d_i=1$ and $\ell_i=1$ ($\epsilon_u=\epsilon_v=3/4$ and $\epsilon_{uv}=0$). The three minima of $f_g$ are $(u_1,v_1)=(0.3087,0.1667)$, $(u_2,v_2)=(0.9060,0.1167)$, and $(u_3,v_3)=(0.4974,0.6667)$.}
\label{fig:xfarplotfgalphabeta}
\end{figure}

Following the variational method introduced in \cite[sec.~II.C]{lin2012mean} (details in \cite[sec.~5.2.3]{lin2012meanthesis}), we obtain the far-field solution for $u$ at the $\alpha \beta$-interface,
\begin{equation} \label{eq:uint12}
u(s)=\frac{u_1+u_2}{2} + \frac{u_1-u_2}{2} \tanh \left[\frac{s}{\delta_{int,\alpha \beta}} \right],
\end{equation}
where we choose $s=0$ as $u=(u_1+u_2)/2$ and define the interfacial width parameter of the $\alpha \beta$-interface as
\begin{equation}\label{eq:intwidth12u}
\delta_{int,\alpha \beta} \equiv \frac{2}{\lvert u_{12} \rvert} \sqrt{\frac{ \hat{\epsilon}_{uv}}{(d_1^2 + d_2^2)}} = \sqrt{2}\sqrt{\frac{\hat{\ell}_1^2 +\hat{\ell}_2^2}{d_1^2 + d_2^2}}.
\end{equation}
Alternatively, if $u_{12}=0$, we can use the relation of $L_3=0$ to obtain $v(s)$ for the $\alpha \beta$-interface
\begin{equation}
v(s)=\frac{v_1+v_2}{2} + \frac{v_1-v_2}{2} \tanh \left[\frac{s}{\delta_{int,\alpha \beta}} \right].
\end{equation}
Fig.~\ref{fig:xfarplotfgalphabeta} illustrates the analytic far-field solutions for the mole fraction $Y_i$ at the $\alpha \beta$-interface for the special case in which all of the weighting coefficient are constants $d_i=1$ and the gradient energy is isotropic $\ell_i= 1$ ($\epsilon_u=\epsilon_v=3/4$ and $\epsilon_{uv}=0$). Compared to the similar analytic far-field solutions for our original potential $f^o$ given by \eqref{eq:potential0} in \cite[fig.~5]{lin2012mean}, the mole fraction $Y_3$ is no longer a constant and, in general, the solutions for $Y_1$ and $Y_2$ are not symmetric with respect to the interface.

The interfacial tension of the $\alpha \beta$-interface in the far-field limit is given by
\begin{equation}\label{eq:gamma12u}
\hat{\sigma}_{\alpha \beta} = \frac{\sqrt{(d_1^2 + d_2^2)\hat{\epsilon}_{uv}}}{3\lvert u_{12} \rvert} B h^4  = \frac{\sqrt{(d_1^2 + d_2^2)(\hat{\ell}_1^2 +\hat{\ell}_2^2)}}{3 \sqrt{2}} B h^4.
\end{equation}
We can calculate the interfacial tensions of the $\beta \gamma $-interface and the $\gamma \alpha$-interface by the same method. Compared to the previous result \eqref{eq:sinthetaintildeell} of the equilibrium dihedral angles for a system with potential $f_g$, a more general relation obeys
\begin{equation}
\begin{split}
\frac{\sin \theta_{\alpha}}{\sqrt{(d_2^2 + d_3^2)(\hat{\ell}_2^2 +\hat{\ell}_3^2)}} &=\frac{\sin \theta_{\beta}}{\sqrt{(d_3^2 + d_1^2)(\hat{\ell}_3^2 +\hat{\ell}_1^2)}} \\
& =\frac{\sin \theta_{\gamma}}{\sqrt{(d_1^2 + d_2^2)(\hat{\ell}_1^2 +\hat{\ell}_2^2)}}.
\end{split}
\end{equation}

\section{Geometry of Interfacial Tension and Line Tension}

Recall the discussion of the mapping from the physical space to the Gibbs space in the Sec.~\ref{sec:mapping}. In a ternary fluid system with three phases, the interfacial tensions associated with the three interfaces correspond to curved lines (trajectories) within the Gibbs triangle, that connect the three minima of a given potential. Each trajectory can be obtained by minimizing the excess grand potential for a two-phase transition in a far-field limit. These trajectories bound a region of the Gibbs space that associates with three-phase transitions. Consequently, the line tension associated with the three-phase contact line is determined by the excess grand potential within this region. This mapping implies a geometrical relationship among the values of interfacial and line tensions and the size and shape of the core area.

Here, we demonstrate this geometrical connection by some special cases of the specific potential $f_g$ given by \eqref{eq:finLIgeneralized}, in which the core area is the small triangle formed by the minima of the potential within the Gibbs triangle.

\subsection{Case: $d_i=1$ and $\ell_i=\ell$}\label{sec:caseequalweightisotropic}

\begin{figure}[h!]
\centering
\includegraphics[scale=0.4]{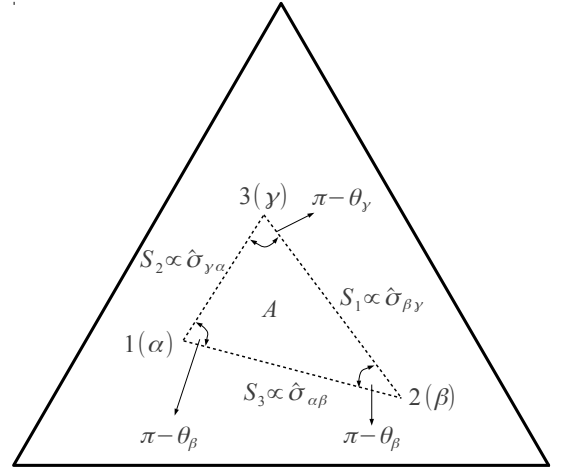}
\caption{\label{fig:gibbsminimasigmageometry} Geometry of the three interfacial tensions in the Gibbs space for a special case of the potential $f_g$ given by \eqref{eq:finLIgeneralized}, in which the weighting coefficients are equal to one $d_i=1$ and the gradient energy is isotropic $\ell_i=1$. The three vertices 1, 2, and 3 correspond to the three minima $\alpha$, $\beta$, and $\gamma$ of $f_g$. The length of the side of the small triangle opposite to the vertex $i$ is labeled by $S_i$, and the area of the small triangle is $A$. The interfacial tension of the $\alpha\beta$-interface $\hat{\sigma}_{\alpha \beta}$ is proportional to the length of the side of the small triangle that connects $\alpha$ and $\beta$, i.e. $S_3$. Similarly, $\hat{\sigma}_{\beta\gamma} \propto S_1$ and $\hat{\sigma}_{\gamma \alpha} \propto S_2$. $\theta_{\alpha}$, $\theta_{\beta}$, and $\theta_{\gamma}$ are the three dihedral angles corresponding to the three-phase contact line.}
\end{figure}

As shown in the Sec.~\ref{sec:straightlinetrajectory}, for the potential $f_g$ in \eqref{eq:finLIgeneralized}, the trajectory for a two-phase transition follows a straight line. In the special case of $f_g$ discussed in Sec.~\ref{sec:dieq1ellieqell},  all of the weighting coefficients $d_i=1$ and the gradient energy is isotropic, $\ell_i=\ell$. Compared to the form \eqref{eq:eqangle} of the classical result, in which the sine of a dihedral angle is proportional to a corresponding interfacial tension, the relation in \eqref{eq:sinthetainsi} indicates that the sine of a dihedral angle is proportional to the length of the side of the small triangle, within the Gibbs triangle, opposite to the vertex for the corresponding bulk phase. This leads to the fact that the three interfacial tensions are proportional to the three lengths of the sides of the small triangle. It also implies that the small triangle is a Neumann triangle, which is similar to the computational boundary of the physical domain (see Fig.~\ref{subfig:realspace} and Fig.~\ref{fig:gibbsminimasigmageometry}).

To prove this property explicitly, we substitute $\hat{\ell}_1^2 +\hat{\ell}_2^2$ from \eqref{eq:gamma12u} into \eqref{eq:twotildeellsqsimple} to obtain 
\begin{equation}
\hat{\sigma}_{\alpha \beta} = \frac{B \ell}{\sqrt{6}} h^2 S_3.
\end{equation}
Similarly, for the $\beta \gamma$-interface and $\gamma \alpha$-interfaces, the tensions are
\begin{equation}
\hat{\sigma}_{\beta \gamma} = \frac{B \ell}{\sqrt{6}} h^2 S_1 \mbox{ and } \hat{\sigma}_{\gamma \alpha} = \frac{B \ell}{\sqrt{6}} h^2 S_2.
\end{equation}

\subsection{Case: $d_i=1$, $\ell_i=\ell$, and $S_i = S$}

We consider a subset of the case treated in Sec.~\ref{sec:caseequalweightisotropic}, in which the small triangle formed by the three minima of the potential $f_g$ in \eqref{eq:finLIgeneralized} is equilateral, i.e. $S_i\equiv S$. Then, the area of the small triangle is
\begin{equation}\label{eq:area}
A=\frac{h}{2}=\frac{\sqrt{3}}{4}S^2.
\end{equation}
From the relation \eqref{eq:twotildeellsqsimple} between the sums of pairs of $\hat{\ell}_i^2$ and $S_i$, we find that
\begin{equation}\label{eq:elli}
\hat{\ell}^2_i \equiv \hat{\ell}^2=\frac{3\ell^2}{2h^4}\frac{S^2}{2}=\frac{3\ell^2}{2h^4} \frac{h}{\sqrt{3}}=\frac{\sqrt{3}\ell^2}{2h^3}.
\end{equation}

\begin{figure}[h!]
\centering
\includegraphics[scale=0.4]{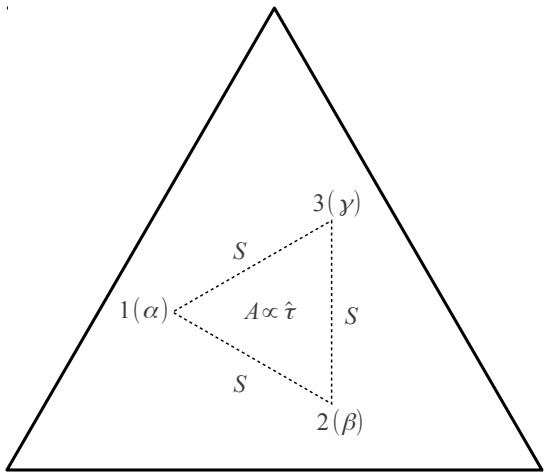}
\caption{\label{fig:gibbsminimataugeometry} Geometry of the line tension in the Gibbs triangle for a special case of $f_g$ with equal weighting coefficients $d_i=1$, isotropic gradient energy $\ell_i=\ell$, and the small triangle formed by the three minima is equilateral. The line tension and the interfacial tensions are proportional to the area $A$ and the lengths of the sides of the small triangle, respectively.}
\end{figure}

According to the form \eqref{eq:omegaxsuv} of the excess grand potential $\hat{\Omega}_{xs}$, we use the Kerins-Boiteux formula \cite{KBpaper} to obtain a line tension associated with the three-phase contact line,
\begin{equation}\label{eq:taufg}
\begin{split}
\hat{\tau} & = B  \int_A \left[ -f_g(u,v) +  g( \nabla u, \nabla v) \right] \mathrm{d}A \\
& = B h^4 \int_A \sum_{i=1}^3 \left[ - Z_i^2(Z_i-1)^2 +  \frac{\hat{\ell}^2}{2} \lvert \nabla Z_i\rvert^2 \right] \mathrm{d}A \\
& = Bh^4\hat{\ell}^2 \tilde{\tau},
\end{split}
\end{equation}
where $\tilde{\tau}$ is the dimensionless quantity introduced in \cite[eq.~53]{lin2012mean}.
Then, from the expressions for area $A$ in \eqref{eq:area} and $\hat{\ell}^2$ in \eqref{eq:elli}, we find
\begin{equation}\label{eq:tauarea}
\hat{\tau} = \sqrt{3} A B\ell^2\tilde{\tau},
\end{equation}
which is proportional to the area of the small triangle, as illustrated in Fig.~\ref{fig:gibbsminimataugeometry}. 

From above cases, the small triangle in the Gibbs space not only tells us the compositions of the three bulk phases in terms of mole fractions from the locations of its three vertices; its size and shape also give us information about the equilibrium dihedral angles and the relative strengths of the interfacial tensions. Moreover, the line tension is proportional to the area of this small triangle.

\section{Summary and Conclusions}

We are interested in potentials that are positive semi-definite that vanish only at three mimima. Moreover, we choose to deal with potentials that vary quadratically around the three minima. We seek potentials in the form of quartic polynomials. Our original potential \cite{lin2012mean} is the sum of three quartic polynomials, each of which vanishes along two parallel lines in the Gibbs triangle. The three minima of our potential are located at the intersections of any three among these six lines. By recognizing this geometrical structure, we can construct generalized quartic potentials with three minima arbitrarily located at the vertices of a small triangle within the Gibbs triangle.

As a first generalization, for each vertex, we can choose a linear function that vanishes along a line that passes through a pair of vertices and another linear function that vanishes along a parallel line that passes through the remaining vertex. The squares of these two linear functions form paraboloids that vanish at these two lines. We use the product of these two paraboloids to construct one of the three quartic polynomials whose sum is the generalized potential. By performing similar constructions at the other two vertex pairs, we can build a potential which is the sum of three quartic polynomials that vanishes at only three points. Our original potential belongs to a subset of this generalization where the two ``vanishing lines'' of each quartic polynomials are parallel. To construct a potential in this subset, we need six parameters, namely the coordinates of the three minima and three weighting coefficients for each quartic polynomials. However, we find that each paraboloid that vanishes at a line passing through one vertex can be generalized to another paraboloid that only vanishes at a point, the vertex. By means of this second generalization, we construct another potential with nine parameters related to the shapes of the new paraboloids, and containing six parameters to characterize  the three minima. We check the properties of these generalized potentials by studying their first and second derivatives.

Generalized potentials in which the valleys connecting any of the two minima follow straight lines in the Gibbs triangle have simple analytic far-field solutions. We prove that this subset of generalized potentials is a special case of our first generalization that was constructed by three pairs of parallel lines. Analytic far-field solutions of this subset are presented. Moreover, we find that this subset is a linear transformation of our original potential. By scaling, we can relate their solutions to those for our original potential. When the weighting coefficients of the potentials in this subset are equal and the gradient energy is isotropic, the lengths of the sides of the small triangle formed by the three minima are proportional to the corresponding interfacial tensions. For the case of equal interfacial tensions, we are able to calculate a line tension that is proportional to the area of the small triangle.

\begin{acknowledgments}
We appreciate the resources and the financial support for this work from the Department of Physics, Carnegie Mellon University. Financial support for C.-Y. Lin by KU Leuven Grant OT/11/063 and for Michael Widom from ONR-MURI under grant N00014-11-1-0678 is gratefully acknowledged. 

\end{acknowledgments}


\appendix

\section{}\label{sec:appendixA}

\begin{figure}[h!]
\centering
\subfigure[\; Geometry of the three minima of $f_1$]{
\label{subfig:gibbsminimaf1}
\includegraphics[scale=0.32]{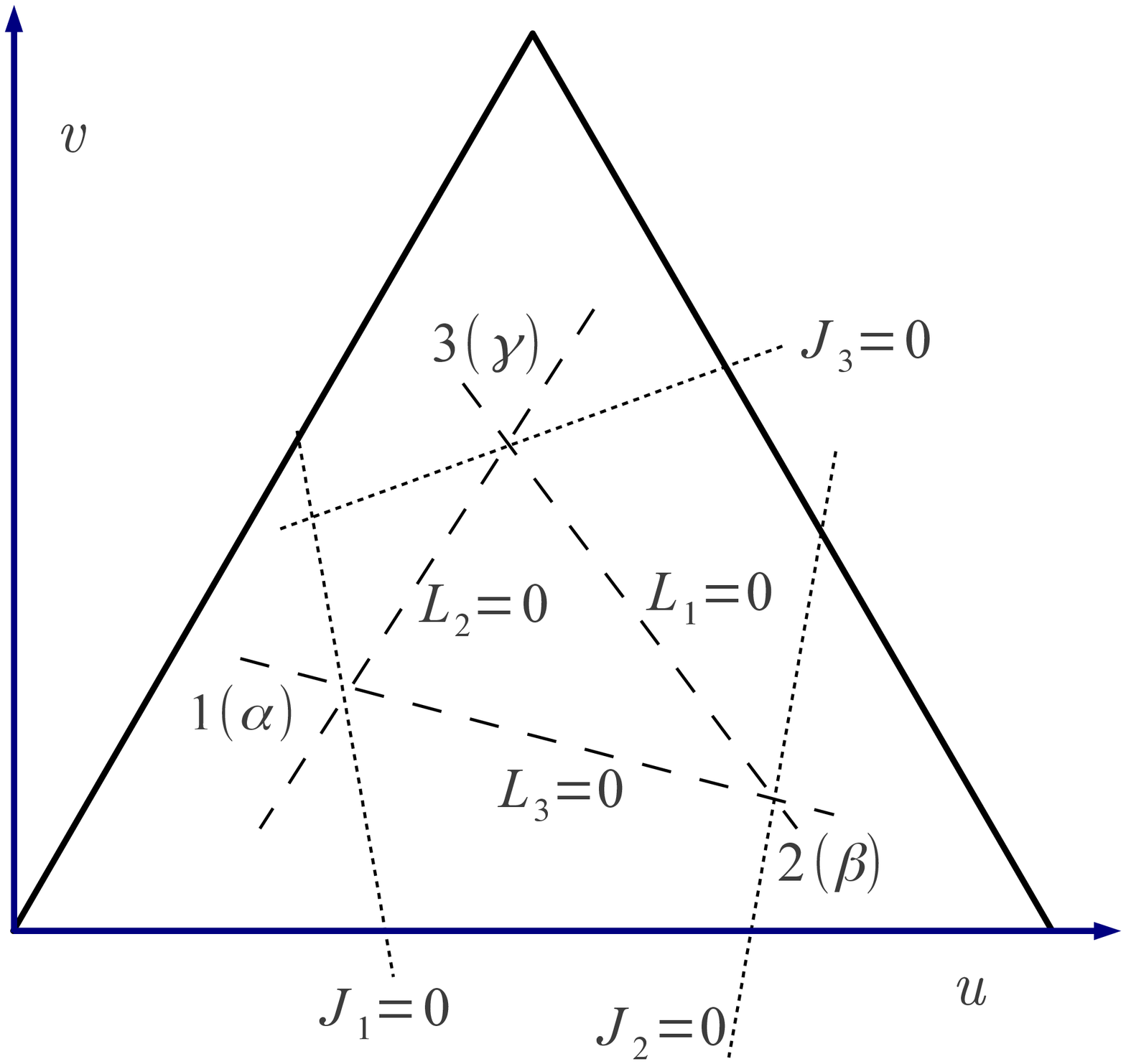}}
\subfigure[\; Contours of $f_1$]{
\label{subfig:gibbsminimaf1sample}
\includegraphics[scale=0.48]{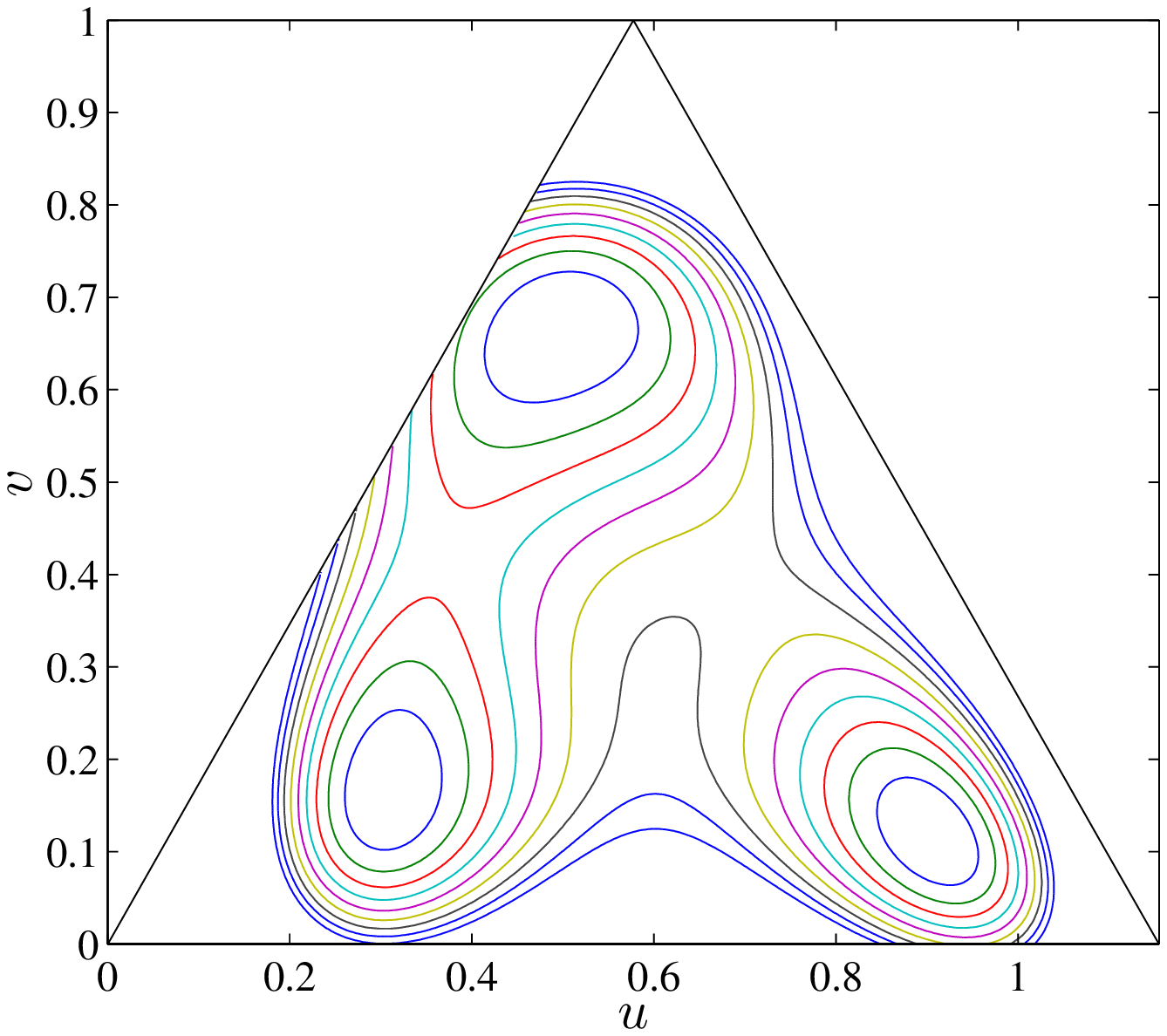}}
\caption{\subref{subfig:gibbsminimaf1} Geometry of the three minima of the quartic potential $f_1$ given by \eqref{eq:f1}. $\alpha=(u_1,v_1)$, $\beta=(u_2,v_2)$, and $\gamma=(u_3,v_3)$ are the three minima of this potential arbitrarily located at any three points in Gibbs space. $L_i=0$ represents a line that passes through $(u_j,v_j)$ and $(u_k,v_k)$ for $j \neq k \neq i$. \subref{subfig:gibbsminimaf1sample} Contour plot of potential $f_1$ for $(u_1,v_1)$ $=$ $(0.3087,0.1667)$, $(u_1,v_1)$ $=$ $(0.3087,0.1667)$, $(u_2,v_2)$ $=$ $(0.9060,0.1167)$, $(u_3,v_3)$ $=$ $(0.4974,0.6667)$, and $(a_1,b_1)=(1,0.1)$, $(a_2,b_2)=(-1,0.1)$, and $(a_3,b_3)=(-0.1,1)$.}
\end{figure}

We can generalize the potential $f_g$ in \eqref{eq:finLIgeneralized} by replacing $I_i$ with an arbitrary linear function $J_i$ that passes through $(u_i,v_i)$. If $J_i(u,v) \equiv a_i (u-u_i) + b_i (v-v_i)$, where $a_i$ and $b_i$ are arbitrary constants, we can obtain a generalized potential
\begin{equation}\label{eq:f1}
f_1=\sum\limits_{i=1}^3 L_i(u,v)^2 J_i(u,v)^2,
\end{equation}
as illustrated in Fig.~\ref{subfig:gibbsminimaf1}. In this case, $J_i=0$ and $L_i=0$ do not need to be parallel.

\begin{figure}[h!]
\centering
\subfigure[\; Geometry of the three minima of $f_2$]{
\label{subfig:gibbsminimaf2}
\includegraphics[scale=0.32]{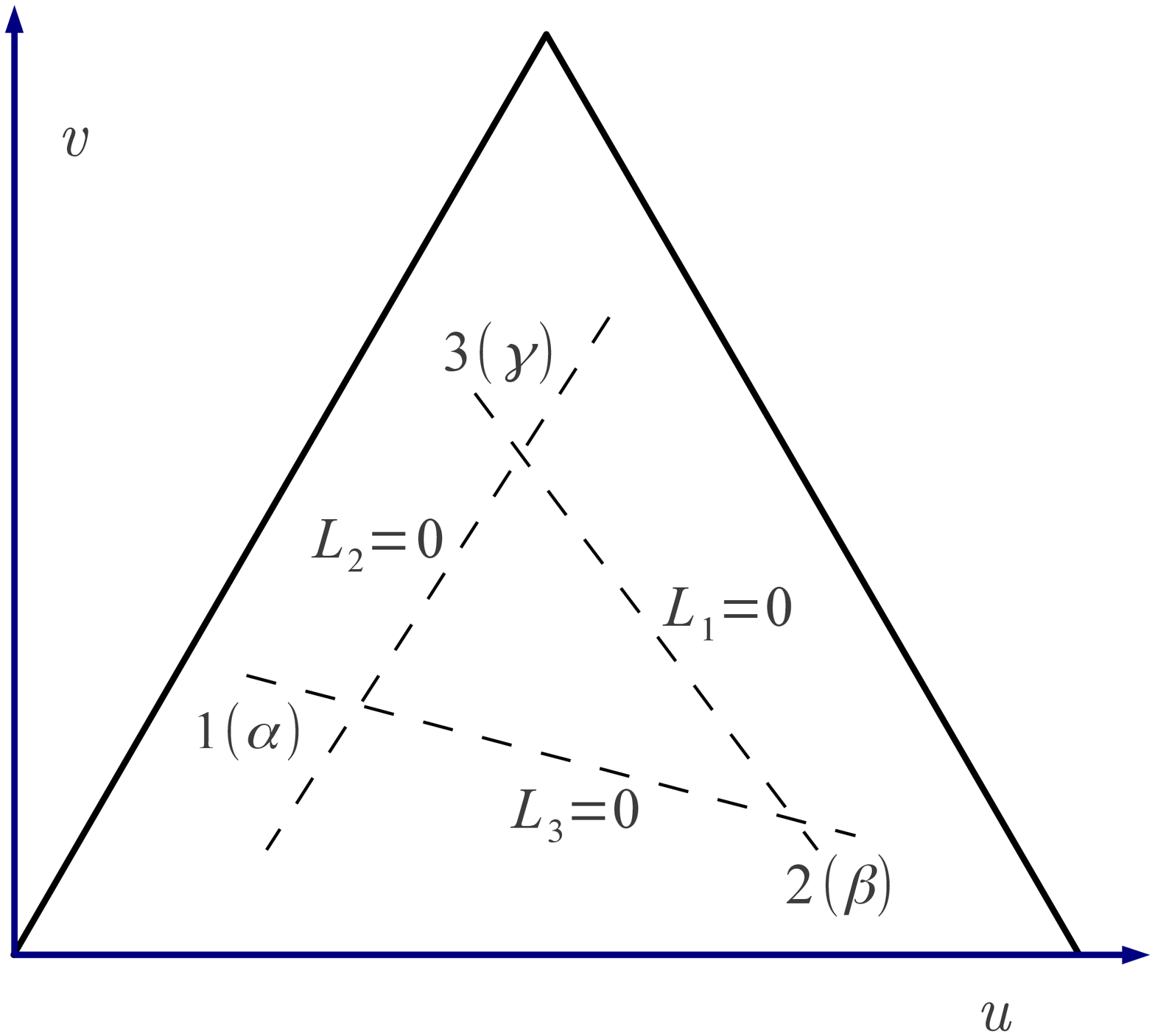}}
\subfigure[\; Contours of $f_2$]{
\label{subfig:gibbsminimaf2sample}
\includegraphics[scale=0.48]{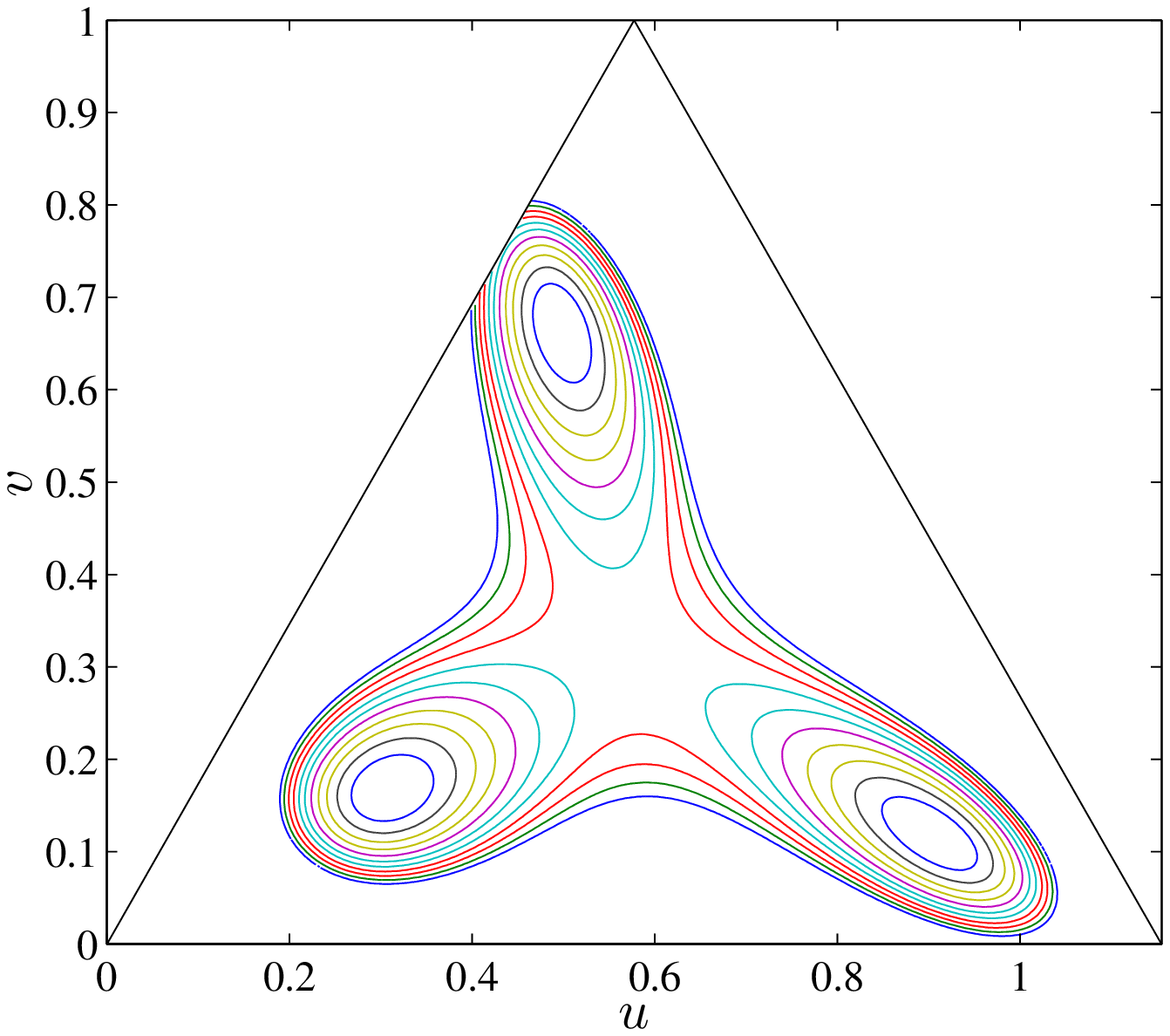}}
\caption{\subref{subfig:gibbsminimaf2} Geometry of the three minima of the quartic potential $f_2$ given by \eqref{eq:f2}. $\alpha=(u_1,v_1)$, $\beta=(u_2,v_2)$, and $\gamma=(u_3,v_3)$ are the three minima of this potential arbitrarily located at any three points in Gibbs space. $L_i=0$ represents a line that passes through $(u_j,v_j)$ and $(u_k,v_k)$ for $j \neq k \neq i$. \subref{subfig:gibbsminimaf2sample} Contour plot of potential $f_2$ for $(u_1,v_1)$ $=$ $(0.3087,0.1667)$, $(u_1,v_1)$ $=$ $(0.3087,0.1667)$, $(u_2,v_2)$ $=$ $(0.9060,0.1167)$, $(u_3,v_3)$ $=$ $(0.4974,0.6667)$, and $d_{12}=d_{23}=d_{31}=1$.}
\end{figure}

For the special case in which $J_1$, $J_2$, and $J_3$ match with $L_2$, $L_3$, and $L_1$, we can define a potential that employs only three lines, namely
\begin{equation}\label{eq:f2}
\begin{split}
f_2 = & d_{12}^2 L_1(u,v)^2 L_2(u,v)^2 \\
& + d_{23}^2 L_2(u,v)^2 L_3(u,v)^2 + d_{31}^2 L_3(u,v)^2 L_1(u,v)^2,
\end{split}
\end{equation}
where $d_{12}$, $d_{23}$, and $d_{31}$ are weighting coefficients. At each minimum, only two lines meet as illustrated in Fig.~\ref{subfig:gibbsminimaf2}. The contours of potential $f_2$ are shown in Fig.~\ref{subfig:gibbsminimaf2sample}. 

Note that $f_1$ in \eqref{eq:f1} and $f_2$ in \eqref{eq:f2} are special cases of the potential $f_G$ defined in \eqref{eq:finLK}.

\section{}\label{sec:appendixBformwithanalyticsoln}

To study the general form of the subset of generalized potentials in which any two-phase transition follows a straight line, we explore the transition between two phases by means of a potential $f_{2D}$ that is a quartic function of three mole fractions having three minima $(u_1,v_1)$, $(u_2,v_2)$, and $(u_3,v_3)$ located at the three vertices of the Gibbs triangle. The potential function along any two of the three wells actually reduces to a one-dimensional potential $f_{1D}$. Consider the transition from $(u_1,v_1)$ to $(u_2,v_2)$ that follows a straight line $L_3=0$. We let
\begin{equation}
\left. f_{2D} \right\rvert _{L_3=0} (u,v) = f_{1D}.
\end{equation}
where $f_{1D}$ is a positive quartic potential function having two minima at $(u_1,v_1)$ and $(u_2,v_2)$. Since $f_{1D}$ follows a straight line $L_3=0$ within the Gibbs triangle, we can express $f_{1D}$ as a function of one variable. Then, we let
\begin{equation}
f_{1D}(u) = (u-u_1)(u-u_2)g(u),
\end{equation}
where $g(u)$ is a quadratic function of $u$. By requiring $\partial f_{1D}/ \partial u = 0$ at $u_1$ and $u_2$ and $\partial^2 f_{1D}/ \partial u^2 > 0$, we obtain
\begin{equation}
f_{1D}(u) \propto (u-u_1)^2(u-u_2)^2.
\end{equation}

We can generalize the contribution of $f_{1D}(u)$ to $f_{2D}(u,v)$, which is denoted as $f^{(1)}_{2D}$ by combination of $L_1(u,v)$ and $K_1(u,v)$, i.e.
\begin{equation}
f^{(1)}_{2D} = L_1^2(u,v)K_1(u,v),
\end{equation}
similar to the potential $f_G$ in \eqref{eq:finLK}, which vanishes at the three minima. However, if the transition between $(u_1,v_1)$ and $(u_2,v_2)$ follows the straight line $L_3=0$, we need
\begin{equation}\label{eq:dLKdhatL3eq0}
\left. \frac{\partial L_1^2(u,v)K_1(u,v)}{\partial \hat{L}_3}) \right\rvert_{L_3=0},
\end{equation}
where $\hat{L}_3 \equiv (v_{12} , -u_{12})$ is a normal vector perpendicular to $L_3=0$ in the Gibbs space. Thus, we calculate
\begin{equation}
\begin{split}
 \frac{\partial (L_1^2 K_1)}{\partial \hat{L}_3} & \left.\right\rvert_{L_3=0}
= 2 \left(\frac{h}{v_{12}} \right) (v-v_1)(v-v_2) 
\\
\times  & \left\{ (v_{12})^2  \left[ \left(a_1 \frac{u_{12}}{v_{12}} +b_1 \right)^2 +c_1^2 \right] (v-v_1) \right.\\
& \;\; \left. - u_{12} \left(\frac{h}{v_{12}}\right) \left[ a_1 b_1 \frac{u_{12}}{v_{12}} + b_1^2 +c_1^2 \right] (v-v_2) \right\},
\end{split}
\end{equation}
in which we substitute $(u-u_1)=(u_{12}/v_{12})(v-v_1)$ for $L_3=0$ by assuming $v_{12} \neq 0$. In general, $h \neq 0$, $u_{12} \neq 0$, $(v-v_1) \neq 0$, and $(v-v_2) \neq 0$. The nontrivial solution for \eqref{eq:dLKdhatL3eq0} is $a_1=v_{12}$, $b_1=-u_{12}$, and $c_1=0$. Then, $f^{(1)}_{2D}$ reduces to
\begin{equation}
f^{(1)}_{2D} = L_1(u,v)^2 I_1(u,v)^2.
\end{equation}

Similarly, we can apply this same analysis to the transitions of $(u_2,v_2) \leftrightarrow (u_3,v_3)$ and $(u_3,v_3) \leftrightarrow (u_1,v_1)$. Thus, we conclude that
\begin{equation}
f_{2D} = f_g.
\end{equation}
In Section~\ref{sec:gpqf}, we have shown that the potential $f_g$ has our desired properties. Moreover, the two-phase transitions of $f_g$ follow straight lines in the space of the Gibbs triangle, which allows us to find analytical solutions in the far-field.

\bibliography{capillarity}

\begin{thebibliography}{10}%
\makeatletter
\providecommand \@ifxundefined [1]{%
 \ifx #1\undefined \expandafter \@firstoftwo
 \else \expandafter \@secondoftwo
\fi
}%
\providecommand \@ifnum [1]{%
 \ifnum #1\expandafter \@firstoftwo
 \else \expandafter \@secondoftwo
\fi
}%
\providecommand \enquote [1]{``#1''}%
\providecommand \bibnamefont  [1]{#1}%
\providecommand \bibfnamefont [1]{#1}%
\providecommand \citenamefont [1]{#1}%
\providecommand\href[0]{\@sanitize\@href}%
\providecommand\@href[1]{\endgroup\@@startlink{#1}\endgroup\@@href}%
\providecommand\@@href[1]{#1\@@endlink}%
\providecommand \@sanitize [0]{\begingroup\catcode`\&12\catcode`\#12\relax}%
\@ifxundefined \pdfoutput {\@firstoftwo}{%
 \@ifnum{\z@=\pdfoutput}{\@firstoftwo}{\@secondoftwo}%
}{%
 \providecommand\@@startlink[1]{\leavevmode\special{html:<a href="#1">}}%
 \providecommand\@@endlink[0]{\special{html:</a>}}%
}{%
 \providecommand\@@startlink[1]{%
  \leavevmode
  \pdfstartlink
   attr{/Border[0 0 1 ]/H/I/C[0 1 1]}%
   user{/Subtype/Link/A<</Type/Action/S/URI/URI(#1)>>}%
  \relax
 }%
 \providecommand\@@endlink[0]{\pdfendlink}%
}%
\providecommand \url  [0]{\begingroup\@sanitize \@url }%
\providecommand \@url [1]{\endgroup\@href {#1}{\urlprefix}}%
\providecommand \urlprefix [0]{URL }%
\providecommand \Eprint[0]{\href }%
\@ifxundefined \urlstyle {%
  \providecommand \doi [1]{doi:\discretionary{}{}{}#1}%
}{%
  \providecommand \doi [0]{doi:\discretionary{}{}{}\begingroup
  \urlstyle{rm}\Url }%
}%
\providecommand \doibase [0]{http://dx.doi.org/}%
\providecommand \Doi[1]{\href{\doibase#1}}%
\providecommand \bibAnnote [3]{%
  \BibitemShut{#1}%
  \begin{quotation}\noindent
    \textsc{Key:}\ #2\\\textsc{Annotation:}\ #3%
  \end{quotation}%
}%
\providecommand \bibAnnoteFile [2]{%
  \IfFileExists{#2}{\bibAnnote {#1} {#2} {\input{#2}}}{}%
}%
\providecommand \typeout [0]{\immediate \write \m@ne }%
\providecommand \selectlanguage [0]{\@gobble}%
\providecommand \bibinfo [0]{\@secondoftwo}%
\providecommand \bibfield [0]{\@secondoftwo}%
\providecommand \translation [1]{[#1]}%
\providecommand \BibitemOpen[0]{}%
\providecommand \bibitemStop [0]{}%
\providecommand \bibitemNoStop [0]{.\EOS\space}%
\providecommand \EOS [0]{\spacefactor3000\relax}%
\providecommand \BibitemShut [1]{\csname bibitem#1\endcsname}%
\bibitem{lin2012mean}%
  \BibitemOpen
  \bibfield{author}{%
  \bibinfo {author} {\bibfnamefont{C.-Y.}\ \bibnamefont{Lin}}, \bibinfo
  {author} {\bibfnamefont{M.}~\bibnamefont{Widom}},\ and\ \bibinfo {author}
  {\bibfnamefont{R.~F.}\ \bibnamefont{Sekerka}},\ }%
  \bibfield{journal}{%
  \bibinfo {journal} {Physical Review E}\ }%
  \textbf{\bibinfo {volume} {85}},\ \bibinfo {pages} {011120} (\bibinfo {year}
  {2012})%
  \bibAnnoteFile{NoStop}{lin2012mean}%
\bibitem{gibbs1928collected}%
  \BibitemOpen
  \bibfield{author}{%
  \bibinfo {author} {\bibfnamefont{J.~W.}\ \bibnamefont{Gibbs}},\ }%
  \emph{\bibinfo {title} {The Collected Works of J. Willard Gibbs:
  Thermodynamics}},\ Vol.~\bibinfo {volume} {1}\ (\bibinfo {publisher}
  {Longmans, Green},\ \bibinfo {year} {1928})%
  \bibAnnoteFile{NoStop}{gibbs1928collected}%
\bibitem{rowlinson2002molecular}%
  \BibitemOpen
  \bibfield{author}{%
  \bibinfo {author} {\bibfnamefont{J.~S.}\ \bibnamefont{Rowlinson}}\ and\
  \bibinfo {author} {\bibfnamefont{B.}~\bibnamefont{Widom}},\ }%
  \emph{\bibinfo {title} {Molecular theory of capillarity}}\ (\bibinfo
  {publisher} {Dover},\ \bibinfo {year} {2002})%
  \bibAnnoteFile{NoStop}{rowlinson2002molecular}%
\bibitem{neumann1894}%
  \BibitemOpen
  \bibfield{author}{%
  \bibinfo {author} {\bibfnamefont{F.~E.}\ \bibnamefont{Neumann}},\ }%
  \enquote{\bibinfo {title} {Vorlesungen \"{u}ber die theorie der
  capillarit\"{a}t (ed. a. wangerin)},}\ \ (\bibinfo {publisher} {Teubner,
  Leipzig},\ \bibinfo {year} {1894})\ Chap.\ \bibinfo {chapter} {6 \S 1}, pp.\
  \bibinfo {pages} {161--2}%
  \bibAnnoteFile{NoStop}{neumann1894}%
\bibitem{adamson1997physical}%
  \BibitemOpen
  \bibfield{author}{%
  \bibinfo {author} {\bibfnamefont{A.~W.}\ \bibnamefont{Adamson}}\ and\
  \bibinfo {author} {\bibfnamefont{A.~P.}\ \bibnamefont{Gast}},\ }%
  \emph{\bibinfo {title} {Physical chemistry of surfaces}}\ (\bibinfo
  {publisher} {Wiley-Interscience},\ \bibinfo {year} {1997})%
  \bibAnnoteFile{NoStop}{adamson1997physical}%
\bibitem{gaydos1987dependence}%
  \BibitemOpen
  \bibfield{author}{%
  \bibinfo {author} {\bibfnamefont{J.}~\bibnamefont{Gaydos}}\ and\ \bibinfo
  {author} {\bibfnamefont{A.~W.}\ \bibnamefont{Neumann}},\ }%
  \bibfield{journal}{%
  \bibinfo {journal} {Journal of colloid and interface science}\ }%
  \textbf{\bibinfo {volume} {120}},\ \bibinfo {pages} {76} (\bibinfo {year}
  {1987})%
  \bibAnnoteFile{NoStop}{gaydos1987dependence}%
\bibitem{drelich1993effect}%
  \BibitemOpen
  \bibfield{author}{%
  \bibinfo {author} {\bibfnamefont{J.}~\bibnamefont{Drelich}}, \bibinfo
  {author} {\bibfnamefont{J.~D.}\ \bibnamefont{Miller}},\ and\ \bibinfo
  {author} {\bibfnamefont{J.}~\bibnamefont{Hupka}},\ }%
  \bibfield{journal}{%
  \bibinfo {journal} {Journal of colloid and interface science}\ }%
  \textbf{\bibinfo {volume} {155}},\ \bibinfo {pages} {379} (\bibinfo {year}
  {1993})%
  \bibAnnoteFile{NoStop}{drelich1993effect}%
\bibitem{grunze1999driven}%
  \BibitemOpen
  \bibfield{author}{%
  \bibinfo {author} {\bibfnamefont{M.}~\bibnamefont{Grunze}},\ }%
  \bibfield{journal}{%
  \bibinfo {journal} {Science}\ }%
  \textbf{\bibinfo {volume} {283}},\ \bibinfo {pages} {41} (\bibinfo {year}
  {1999})%
  \bibAnnoteFile{NoStop}{grunze1999driven}%
\bibitem{weigl1999microfluidic}%
  \BibitemOpen
  \bibfield{author}{%
  \bibinfo {author} {\bibfnamefont{B.~H.}\ \bibnamefont{Weigl}}\ and\ \bibinfo
  {author} {\bibfnamefont{P.}~\bibnamefont{Yager}},\ }%
  \bibfield{journal}{%
  \bibinfo {journal} {Science}\ }%
  \textbf{\bibinfo {volume} {283}},\ \bibinfo {pages} {346} (\bibinfo {year}
  {1999})%
  \bibAnnoteFile{NoStop}{weigl1999microfluidic}%
\bibitem{hienola2007estimation}%
  \BibitemOpen
  \bibfield{author}{%
  \bibinfo {author} {\bibfnamefont{A.~I.}\ \bibnamefont{Hienola}}, \bibinfo
  {author} {\bibfnamefont{P.~M.}\ \bibnamefont{Winkler}}, \bibinfo {author}
  {\bibfnamefont{P.~E.}\ \bibnamefont{Wagner}}, \bibinfo {author}
  {\bibfnamefont{H.}~\bibnamefont{Vehkam{\"a}ki}}, \bibinfo {author}
  {\bibfnamefont{A.}~\bibnamefont{Lauri}}, \bibinfo {author}
  {\bibfnamefont{I.}~\bibnamefont{Napari}},\ and\ \bibinfo {author}
  {\bibfnamefont{M.}~\bibnamefont{Kulmala}},\ }%
  \bibfield{journal}{%
  \bibinfo {journal} {The Journal of chemical physics}\ }%
  \textbf{\bibinfo {volume} {126}},\ \bibinfo {pages} {094705} (\bibinfo {year}
  {2007})%
  \bibAnnoteFile{NoStop}{hienola2007estimation}%
\bibitem{sackmann2002cell}%
  \BibitemOpen
  \bibfield{author}{%
  \bibinfo {author} {\bibfnamefont{E.}~\bibnamefont{Sackmann}}\ and\ \bibinfo
  {author} {\bibfnamefont{R.~F.}\ \bibnamefont{Bruinsma}},\ }%
  \bibfield{journal}{%
  \bibinfo {journal} {ChemPhysChem}\ }%
  \textbf{\bibinfo {volume} {3}},\ \bibinfo {pages} {262} (\bibinfo {year}
  {2002})%
  \bibAnnoteFile{NoStop}{sackmann2002cell}%
\bibitem{fukai1995wetting}%
  \BibitemOpen
  \bibfield{author}{%
  \bibinfo {author} {\bibfnamefont{J.}~\bibnamefont{Fukai}}, \bibinfo {author}
  {\bibfnamefont{Y.}~\bibnamefont{Shiiba}}, \bibinfo {author}
  {\bibfnamefont{T.}~\bibnamefont{Yamamoto}}, \bibinfo {author}
  {\bibfnamefont{O.}~\bibnamefont{Miyatake}}, \bibinfo {author}
  {\bibfnamefont{D.}~\bibnamefont{Poulikakos}}, \bibinfo {author}
  {\bibfnamefont{C.~M.}\ \bibnamefont{Megaridis}},\ and\ \bibinfo {author}
  {\bibfnamefont{Z.}~\bibnamefont{Zhao}},\ }%
  \bibfield{journal}{%
  \bibinfo {journal} {Physics of Fluids}\ }%
  \textbf{\bibinfo {volume} {7}},\ \bibinfo {pages} {236} (\bibinfo {year}
  {1995})%
  \bibAnnoteFile{NoStop}{fukai1995wetting}%
\bibitem{bonn2009wetting}%
  \BibitemOpen
  \bibfield{author}{%
  \bibinfo {author} {\bibfnamefont{D.}~\bibnamefont{Bonn}}, \bibinfo {author}
  {\bibfnamefont{J.}~\bibnamefont{Eggers}}, \bibinfo {author}
  {\bibfnamefont{J.}~\bibnamefont{Indekeu}}, \bibinfo {author}
  {\bibfnamefont{J.}~\bibnamefont{Meunier}},\ and\ \bibinfo {author}
  {\bibfnamefont{E.}~\bibnamefont{Rolley}},\ }%
  \bibfield{journal}{%
  \bibinfo {journal} {Reviews of modern physics}\ }%
  \textbf{\bibinfo {volume} {81}},\ \bibinfo {pages} {739} (\bibinfo {year}
  {2009})%
  \bibAnnoteFile{NoStop}{bonn2009wetting}%
\bibitem{indekeu2010wetting}%
  \BibitemOpen
  \bibfield{author}{%
  \bibinfo {author} {\bibfnamefont{J.~O.}\ \bibnamefont{Indekeu}},\ }%
  \bibfield{journal}{%
  \bibinfo {journal} {Physica A: Statistical Mechanics and its Applications}\
  }%
  \textbf{\bibinfo {volume} {389}},\ \bibinfo {pages} {4332} (\bibinfo {year}
  {2010})%
  \bibAnnoteFile{NoStop}{indekeu2010wetting}%
\bibitem{amirfazli2004status}%
  \BibitemOpen
  \bibfield{author}{%
  \bibinfo {author} {\bibfnamefont{A.}~\bibnamefont{Amirfazli}}\ and\ \bibinfo
  {author} {\bibfnamefont{A.~W.}\ \bibnamefont{Neumann}},\ }%
  \bibfield{journal}{%
  \bibinfo {journal} {Advances in colloid and interface science}\ }%
  \textbf{\bibinfo {volume} {110}},\ \bibinfo {pages} {121} (\bibinfo {year}
  {2004})%
  \bibAnnoteFile{NoStop}{amirfazli2004status}%
\bibitem{schimmele2007conceptual}%
  \BibitemOpen
  \bibfield{author}{%
  \bibinfo {author} {\bibfnamefont{L.}~\bibnamefont{Schimmele}}, \bibinfo
  {author} {\bibfnamefont{M.}~\bibnamefont{Napi{\'o}rkowski}},\ and\ \bibinfo
  {author} {\bibfnamefont{S.}~\bibnamefont{Dietrich}},\ }%
  \bibfield{journal}{%
  \bibinfo {journal} {The Journal of chemical physics}\ }%
  \textbf{\bibinfo {volume} {127}},\ \bibinfo {pages} {164715} (\bibinfo {year}
  {2007})%
  \bibAnnoteFile{NoStop}{schimmele2007conceptual}%
\bibitem{anderson1998diffuse}%
  \BibitemOpen
  \bibfield{author}{%
  \bibinfo {author} {\bibfnamefont{D.~M.}\ \bibnamefont{Anderson}}, \bibinfo
  {author} {\bibfnamefont{G.~B.}\ \bibnamefont{McFadden}},\ and\ \bibinfo
  {author} {\bibfnamefont{A.~A.}\ \bibnamefont{Wheeler}},\ }%
  \bibfield{journal}{%
  \bibinfo {journal} {Annual review of fluid mechanics}\ }%
  \textbf{\bibinfo {volume} {30}},\ \bibinfo {pages} {139} (\bibinfo {year}
  {1998})%
  \bibAnnoteFile{NoStop}{anderson1998diffuse}%
\bibitem{LandauLifshitz1935}%
  \BibitemOpen
  \bibfield{author}{%
  \bibinfo {author} {\bibfnamefont{L.~D.}\ \bibnamefont{Landau}}\ and\ \bibinfo
  {author} {\bibfnamefont{E.~M.}\ \bibnamefont{Lifshitz}},\ }%
  \bibfield{journal}{%
  \bibinfo {journal} {Phys. Zeit. Sowjetunion}\ }%
  \textbf{\bibinfo {volume} {8}} (\bibinfo {year} {1935}),\ \bibinfo {note}
  {see also, "Electrodynamics of Continuous Media",\S45, p.158, reprinted by
  Beijing World Publishing Cooperation by arrangement with
  Butterworth-Heinemann (1999)}%
  \bibAnnoteFile{NoStop}{LandauLifshitz1935}%
\bibitem{umantsev2012field}%
  \BibitemOpen
  \bibfield{author}{%
  \bibinfo {author} {\bibfnamefont{A.}~\bibnamefont{Umantsev}},\ }%
  \emph{\bibinfo {title} {Field Theoretic Method in Phase Transformations}},\
  Vol.\ \bibinfo {volume} {840}\ (\bibinfo {publisher} {Springer New York},\
  \bibinfo {year} {2012})%
  \bibAnnoteFile{NoStop}{umantsev2012field}%
\bibitem{szleifer1992surface}%
  \BibitemOpen
  \bibfield{author}{%
  \bibinfo {author} {\bibfnamefont{I.}~\bibnamefont{Szleifer}}\ and\ \bibinfo
  {author} {\bibfnamefont{B.}~\bibnamefont{Widom}},\ }%
  \bibfield{journal}{%
  \bibinfo {journal} {Mol. Phys.}\ }%
  \textbf{\bibinfo {volume} {75}},\ \bibinfo {pages} {925} (\bibinfo {year}
  {1992})%
  \bibAnnoteFile{NoStop}{szleifer1992surface}%
\bibitem{koga2010first}%
  \BibitemOpen
  \bibfield{author}{%
  \bibinfo {author} {\bibfnamefont{K.}~\bibnamefont{Koga}}, \bibinfo {author}
  {\bibfnamefont{J.~O.}\ \bibnamefont{Indekeu}},\ and\ \bibinfo {author}
  {\bibfnamefont{B.}~\bibnamefont{Widom}},\ }%
  \bibfield{journal}{%
  \bibinfo {journal} {Faraday Discuss.}\ }%
  \textbf{\bibinfo {volume} {146}},\ \bibinfo {pages} {217} (\bibinfo {year}
  {2010})%
  \bibAnnoteFile{NoStop}{koga2010first}%
\bibitem{koga2008mean}%
  \BibitemOpen
  \bibfield{author}{%
  \bibinfo {author} {\bibfnamefont{K.}~\bibnamefont{Koga}}\ and\ \bibinfo
  {author} {\bibfnamefont{B.}~\bibnamefont{Widom}},\ }%
  \bibfield{journal}{%
  \bibinfo {journal} {J. Chem. Phys.}\ }%
  \textbf{\bibinfo {volume} {128}},\ \bibinfo {pages} {114716} (\bibinfo {year}
  {2008})%
  \bibAnnoteFile{NoStop}{koga2008mean}%
\bibitem{koga2010infinite}%
  \BibitemOpen
  \bibfield{author}{%
  \bibinfo {author} {\bibfnamefont{K.}~\bibnamefont{Koga}}, \bibinfo {author}
  {\bibfnamefont{J.~O.}\ \bibnamefont{Indekeu}},\ and\ \bibinfo {author}
  {\bibfnamefont{B.}~\bibnamefont{Widom}},\ }%
  \bibfield{journal}{%
  \bibinfo {journal} {Physical review letters}\ }%
  \textbf{\bibinfo {volume} {104}},\ \bibinfo {pages} {36101} (\bibinfo {year}
  {2010})%
  \bibAnnoteFile{NoStop}{koga2010infinite}%
\bibitem{lin2012meanthesis}%
  \BibitemOpen
  \bibfield{author}{%
  \bibinfo {author} {\bibfnamefont{C.}~\bibnamefont{Lin}},\ }%
  \emph{\bibinfo {title} {Mean-field density functional theory of a three-phase
  contact line}}\ (\bibinfo {publisher} {PhD thesis, Carnegie Mellon
  University},\ \bibinfo {year} {2012})%
  \bibAnnoteFile{NoStop}{lin2012meanthesis}%
\bibitem{varea1992statistical}%
  \BibitemOpen
  \bibfield{author}{%
  \bibinfo {author} {\bibfnamefont{C.}~\bibnamefont{Varea}}\ and\ \bibinfo
  {author} {\bibfnamefont{A.}~\bibnamefont{Robledo}},\ }%
  \bibfield{journal}{%
  \bibinfo {journal} {Physica A: Statistical and Theoretical Physics}\ }%
  \textbf{\bibinfo {volume} {183}},\ \bibinfo {pages} {12} (\bibinfo {year}
  {1992})%
  \bibAnnoteFile{NoStop}{varea1992statistical}%
\bibitem{KBpaper}%
  \BibitemOpen
  \bibfield{author}{%
  \bibinfo {author} {\bibfnamefont{J.}~\bibnamefont{Kerins}}\ and\ \bibinfo
  {author} {\bibfnamefont{M.}~\bibnamefont{Boiteux}},\ }%
  \bibfield{journal}{%
  \bibinfo {journal} {Physica A}\ }%
  \textbf{\bibinfo {volume} {117}},\ \bibinfo {pages} {575} (\bibinfo {year}
  {1983})%
  \bibAnnoteFile{NoStop}{KBpaper}%
\end{thebibliography}%
\bibliographystyle{apsrev4-1}

\end{document}